\documentclass[onecolumn,amsmath,amssymb,12pt,superscriptaddress,nofootinbib]{revtex4}
\pdfoutput=1

\usepackage[latin1]{inputenc}
\usepackage[english]{babel}
\usepackage{amssymb}
\usepackage{amsmath}
\usepackage{amsthm}
\usepackage{float}
\usepackage[]{graphicx}
\usepackage[]{subfigure}
\usepackage{tensor}
\usepackage{color}
\usepackage{cancel}
\usepackage{setspace}
\usepackage{fancyhdr}
\usepackage{framed}
\usepackage{braket}
\usepackage{siunitx}
\usepackage{tikz}
\newcommand{\be}{\begin{equation}}
\newcommand{\ee}{\end{equation}}
\usepackage[bookmarks,linktocpage, colorlinks=true, plainpages = false, citecolor = blue,  linkcolor=blue, urlcolor = magenta, filecolor = blue]{hyperref} 

\usepackage{mathrsfs}
\usepackage[title]{appendix}
\numberwithin{equation}{section}

\begin{document}

\allowdisplaybreaks

\title{Prospects for testing the inverse-square law and gravitomagnetism using quantum interference}
\vspace{.3in}

\author{Fay\c{c}al Hammad}
\email{fhammad@ubishops.ca}
\affiliation{Department of Physics and Astronomy, Bishop's University, 2600 College Street, Sherbrooke, QC, J1M~1Z7
Canada}
\affiliation{Physics Department, Champlain 
College-Lennoxville, 2580 College Street, Sherbrooke,  
QC, J1M~0C8 Canada}
\affiliation{D\'epartement de Physique, Universit\'e de Montr\'eal,\\
2900 Boulevard \'Edouard-Montpetit,
Montr\'eal, QC, H3T 1J4
Canada} 

\author{Alexandre Landry} \email{alexandre.landry.1@umontreal.ca} 
\affiliation{D\'epartement de Physique, Universit\'e de Montr\'eal,\\
2900 Boulevard \'Edouard-Montpetit,
Montr\'eal, QC, H3T 1J4
Canada} 

\author{Kaleb Mathieu} \email{kmathieu17@ubishops.ca} 
\affiliation{Department of Physics and Astronomy, Bishop's University, 2600 College Street, Sherbrooke, QC, J1M~1Z7
Canada}

\begin{abstract}
We examine a simple tabletop experimental setup for probing the inverse-square law of gravity and detecting eventual deviations therefrom. The nature of the setup allows indeed to effectively reach for shorter distances compared to what is allowed by other methods. Furthermore, we show that the same setup could also in principle be used to probe the interaction between gravitomagnetism and the intrinsic angular spin of quantum particles. Moreover, we show that the setup allows to have a gravitationally induced harmonic oscillator, introducing thus the possibility of studying in a novel way the interaction between gravity and quantum particles.
\end{abstract}


\maketitle

\section{Introduction}\label{sec:Intro}
For large masses and high velocities, General Relativity (GR) ---the theory that currently provides the best description of gravity--- successfully provides well-tested relativistic corrections to Newton's law of gravitation at long separation distances between the interacting masses \cite{Will2018}. However, in the weak-field regime, GR simply reduces to Newtonian gravity no matter what the separation distance is between the two masses. In fact, GR does not suggest any deviation from the inverse-square law (ISL) for the gravitational interaction, $Gm_1m_2/r^2$, where $G$ is Newton's constant and $r$ is the distance between two static masses $m_1$ and $m_2$ \cite{Will2018,Will2016}.

In contrast to GR, many models aiming at modifying gravity \cite{ModifiedGravity2019} or at unifying gravity with the other three fundamental forces of Nature predict some form of deviation from this ISL \cite{NonNewtonianBook,Problems,LongPrice2003,Tests2003,TestsReview2014}. Three main classes of departures from the ISL have been intensively investigated in the literature. The most favored one is a formula usually given in terms of the gravitational interaction energy $U(r)$ between the two masses, displaying a Yukawa-like correction:
\begin{equation}\label{Yukawa}
U(r)= -\frac{Gm_1m_2}{r}\left(1+\alpha e^{-r/\lambda}\right).
\end{equation}
The first term in this formula represents the Newtonian potential while the second term is the correction. The constant $\alpha$ is dimensionless and it quantifies the strength of the deviation from the ISL, i.e., the strength of a possible ``fifth force''. Such a constant might in principle depend on the baryonic composition of the test masses in the case of particle physics models, in which case the Weak Equivalence Principle (WEP) of GR becomes also violated. On the other hand, within standard spacetime-based gravitational theories, like in $f(R)$-modified gravitational theories, the parameters $\alpha$ and $\lambda$ arise in the post-Newtonian limit of the theory: $\alpha$ depends on the first derivative of the modified gravitational Lagrangian $f(R)$ with respect to the Ricci scalar $R$, whereas the parameter $\lambda$ depends on the first and second derivatives of the functional $f(R)$ (see, e.g., Refs.\,\cite{f(R)1,f(R)2}). The parameter $\lambda$ has the dimensions of a length and it quantifies the range of the fifth force. It is the smallness of the parameter $\lambda$, appearing in the denominator of a decaying exponential, that makes it difficult to experimentally bring into evidence this minute possible correction to the ISL. The correction, if any, is thus exponentially suppressed with the separation distance. In fact, the parameter $\lambda$ is constrained (for certain values of $\alpha$) by various short-range gravity experiments ---that continues to be improved--- to not exceed $\sim10\,\mu{\rm m}$ \cite{Tests1985,Tests1993,TestsReview1999,Tests2001,Tests2002,Tests2003,Tests2007,Tests2009,Tests2011,Tests2012,TestsReview2014,Tests2015,Tests2016,Tests2018}. To detect such a deviation one needs thus to probe separation distances of the order of the micrometer. Obviously, electromagnetic, van der Waals and Casimir forces all make it the more difficult to attain the desired precision to distinguish these other forces from a purely gravitational contribution based on tabletop experiments. A similar issue arises at particle accelerators, where very short distances are probed, because of the weakness of gravity compared to the other three fundamental forces. We refer, however, to the recent proposal in Ref.~\cite{GravityLandau} for a new way of putting into evidence deviations from the ISL based on the effect of gravity on the quantized Landau levels (see also Refs.\,\cite{QHE,Josephson, GravityLandauII} for recent prospects for exploiting other quantum phenomena to reach such a goal).  

Two other classes of deviations from the ISL are the so-called power-law potentials. These are also well investigated \cite{Problems,LongPrice2003,Tests2003,TestsReview2014}, although less favored and less tested by the community compared to the Yukawa-like deviation \cite{TestsReview2014}. The first of these classes gives a gravitational potential energy between two masses in the form,
\begin{equation}\label{PL}
U(r)= -\frac{Gm_1m_2}{r}\left[1+\left(\frac{r_0}{r}\right)^{n}\right].
\end{equation}
This class of potentials arises mainly ---but not only--- within the extensions of the Standard Model of particle physics and models of spacetime involving extra spatial dimensions. We have chosen here, for practical purposes, the simple form (\ref{PL}) given in terms of a single parameter $r_0$ that has the dimensions of a length \cite{TestsReview2014}. This parameter quantifies therefore both the strength of the fifth force and the range of the latter. It is model-dependent, and hence composition-dependent, which means that it might also in principle lead to violations of the WEP. The power $n$ is an integer. In the extra spatial dimensions scenarios,  it is usually taken to be $n\leq6$ and it is assumed to take on a fixed value for a specific number $n$ of extra dimensions. In this case, the factor $r_0^n$ becomes a composition-independent parameter.

The second class of power-law potentials frequently found in the literature arises exclusively from models involving extra dimensions of space and takes the form,
\begin{equation}\label{PLADD}
U(r)= -\frac{G_nm_1m_2}{r^{n+1}}.
\end{equation}
The constant $G_n$ denotes a new gravitational constant, which depends on the number $n$ of the extra spatial dimensions of spacetime. It reduces to Newton's constant in the low-energy limit. Formula (\ref{PLADD}) is supposed to be valid for distances below the scale of the extra dimensions. Therefore, this expression is usually recast in the more general form (\ref{PL}) for convenience \cite{TestsReview2014}. For this reason, we are going to consider in this paper only formula (\ref{PL}) for the case of power-law deviations from the ISL.

It is clear that the worse problem one faces when attempting to use massive gravitational sources for testing these various possible deviations from the ISL is the fact that only the closest layers of the massive source would efficiently participate in the interaction \cite{Problems}. Our goal in this paper is therefore to examine a setup that would not only be able to attain the required short distances and be able to avoid the other non-gravitational interactions \cite{Constraints2004,Constraints2005,YukawaIJMP2011,f(R)Yukawa,Saha2013,NewConstraints}, but, above all, be also able to ``use up'' the whole mass source. The strategy is thus to get the test particle not only closer to the gravitational source, but literally delve deep inside the latter and interact with every layer of the source all the way to the heart of the latter. As we shall see, the very configuration of our setup renders the latter indeed capable of realizing each one of these objectives and thus test the various possible deviations from the ISL all at once.

Now, it is also well known that in the case of a rotating massive source, GR predicts deviations from Newtonian gravity thanks to the so-called frame-dragging effect which, in turn, gives rise to the so-called gravitomagnetism \cite{CiufoliniWheeler,Pfister,AboutGM,GMinQM,GMinf(R)}. It is believed that intrinsic spin of quantum particles could couple to this gravitomagnetism and hence provide a means for testing the latter. However, the problem one faces when trying to test gravitomagnetism is again, as discussed by various authors \cite{Biemond,Ciufolini,Gravitomagnetism1999,Gravitomagnetism2000,Tajmar,Ahmedov,Gravitomagnetism2018}, the weakness of such an interaction\footnote{See Ref.~\cite{GravityLandauII} for a new proposal for testing the frame-dragging effect using quantum particles.}. As we shall see, although it cannot presently improve much on the precision one can reach in measuring gravitomagnetism, our setup does constitute a means that is way more efficient and more suitable for fully testing gravitomagnetism once effective noise elimination techniques and higher phase-shifts detection precision are achieved.

Finally, the same setup, as we shall see, is also suitable for  studying the quantum behavior of particles within a classical gravitational field. The configuration of the massive source used in our setup is indeed capable of gravitationally inducing a simple harmonic oscillation in the motion of a quantum particle. The energy levels of the particle become thus quantized by the gravitational field in a way similar to what is done based on the behavior of cold neutrons inside the Earth's gravitational field \cite{LectureNeutronsGravity,QBounce,QuBounce,RoleofNeutrons,Abele,LorentzViolation2019}. Therefore, any deviation from the quantized energy levels caused by Newtonian gravity would automatically indicate a departure from the ISL. 

The rest of this paper is organized as follows. In the next section we describe our setup and explain its working principle. In Section \ref{sec:III}, we compute the phase shift that would be recorded by such a setup as it is induced on the quantum particles used in the experiment for each of the two ISL deviation classes (\ref{Yukawa}) and (\ref{PL}). In Section \ref{sec:IV}, we compute the phase shift that would result when the test particles interact with the gravitomagnetism of the massive source used in the setup. In Section \ref{sec:V}, we explain how one could use the setup to create a gravitationally induced harmonic oscillator. We find the quantized energy levels of the latter caused by the Newtonian potential and then compute the correction brought to such levels by deviations from the ISL. We conclude this paper with a short Summary \& Discussion Section.

\section{The setup}\label{sec:II}
\begin{figure}[H]
    \centering
    \includegraphics[scale=0.6]{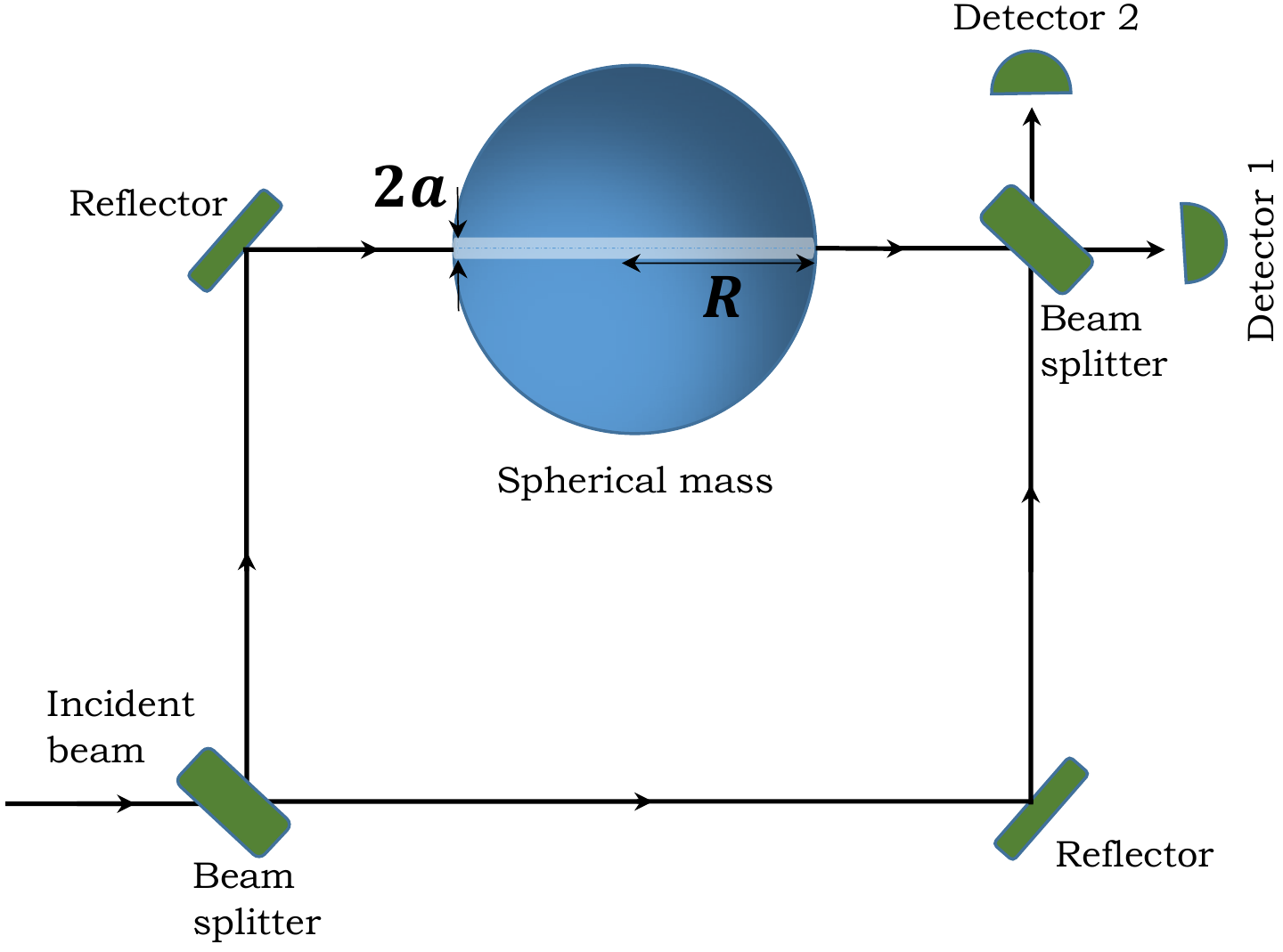}
    \caption{A Mach-Zehnder interferometer with a massive sphere, containing a narrow tunnel drilled across its diameter and lying along one of the arms of the interferometer.}
    \label{COWBall}
\end{figure}

Our setup consists simply of a massive homogeneous solid sphere, inside of which a narrow cylindrical tunnel of radius $a$ is drilled across the diameter of the sphere to make a pathway for the neutrons, atoms, or even the molecules, used in the experiment. Such a special configuration of the mass source allows, as mentioned in the Introduction, to use up every layer of the mass. The sphere is to be placed in one of the arms of a Mach-Zehnder interferometer as shown in figure \ref{COWBall}. 
As we shall see later, the diameter $2a$ of the tunnel inside the sphere should be made small enough compared to the radius of the sphere in order to be able to extract useful approximations for the resulting phase shifts as well as to be able to gravitationally induce a harmonic oscillator. However, in order to avoid unwanted diffraction of the beam of particles, the diameter $2a$ of the tunnel should be made much larger than the Compton wavelength $\lambda_C$ of the test particles used in the experiment. 

The interferometer detects the difference in the quantum phase between the two beams going through its two arms in a way very similar to a COW experiment \cite{COWReview}. Indeed, the number of registered clicks at each of the two detectors ---located after the second beam splitter on the right--- depends on the relative phase between the two beams. 

The beam going through the solid sphere in the positive $x$-direction experiences one of the gravitational potentials leading to either energies (\ref{Yukawa}) or (\ref{PL}). As a consequence, that beam acquires, due to the change $\Delta p(x)$ in the linear momentum $p(x)$ of the particles, like in a COW experiment, an extra phase of $\Delta\phi=(2/\hbar)\int_{0}^{R}\Delta p(x)\,{\rm d}x$ relative to the reference beam which did not go through the massive sphere. For prepared wave packets of de Broglie's wavelength $\lambda_0$ and wave number $k_{0}=p_{0}/\hbar$, the energy of the particles of mass $m$ is $\mathcal{E}_0=\hbar^2k_{0}^2/2m\gg U(x)$. That is, the gravitational interaction of the particles with the massive sphere is treated as a small perturbation. The conservation of energy of the particles of mass $m$ then leads to $\mathcal{E}_0=p^2(x)/2m+U(x)$, so that $\Delta p(x)=p-p_0\approx-p_0U(x)/2\mathcal{E}_0=-m\lambda_0U(x)/h$. Thus, the phase shift between the two beams is simply given by $\Delta\phi=-(m\lambda_0/\pi\hbar^2)\int_{0}^{R}U(x)\,{\rm d}x$. Therefore, substituting into this formula successively the two potential energies (\ref{Yukawa}) and (\ref{PL}) would give us at once the different phase shifts to be expected to result between the two beams of the interferometer.
\section{Phase shifts due to deviations from the ISL}\label{sec:III}
In this section we are going then to compute the phase shift due to each of the above two forms of deviations from the ISL. For that purpose, we need to compute the gravitational potential felt by the particle at any distance $x$ from the center of the sphere based on each of the formulas (\ref{Yukawa}) and (\ref{PL}). To properly take into account the presence of the cylindrical tunnel inside the sphere, we first need to compute both the gravitational potential inside a full sphere of radius $R$ at the distance $x$ from its center and the one due to a uniform full cylinder of radius $a$ and length $2R$ at a distance $x$ from its middle point along its axis of symmetry. Then, we simply subtract the potential due to the cylinder from the one due to the sphere to find the effective potential felt by the particle traveling through the tunnel. We are going to follow this strategy for each of the two possible deviations (\ref{Yukawa}) and (\ref{PL}).

\subsection{With a Yukawa-like deviation}\label{sec:IIIA}
First, we easily see that formula (\ref{Yukawa}) gives the gravitational potential at a distance $x$ along the axis of symmetry of a ring of uniform linear mass density $\mu$ and radius $y$ in the form $-2\pi G\mu\,y(x^2+y^2)^{-\frac{1}{2}}(1+\alpha e^{-\sqrt{x^2+y^2}/\lambda})$. By integrating this expression over the concentric rings that make up a uniform disk, we easily deduce also the gravitational potential at a distance $x$ along the axis of symmetry of a massive disk of radius $y$ and of uniform surface mass density $\sigma$. We find, $-2\pi G\sigma[\sqrt{x^2+y^2}-x+\alpha\lambda(e^{-x/\lambda}- e^{-\sqrt{x^2+y^2}/\lambda})]$. 

Using this last expression for the uniform massive disk, we easily also compute the gravitational potential along the axis of symmetry of a full cylinder of radius $a$, of uniform density $\rho$ and of length $2R$, at a distance $x$ from its middle point. It suffices indeed to integrate the potential due to a uniform disk over the continuous distribution of disks that make up the cylinder. The detailed calculation is given in \ref{A} and the result is $V_C(x)$ as given by Eq.~(\ref{CylinderYuk}). Similarly, by integrating over the continuous distribution of uniform disks that make up a sphere, we find the gravitational potential inside a sphere of radius $R$ and of uniform mass density $\rho$ at any distance $x$ from its center. The result is $V_S(x)$ as given by Eq.~(\ref{SphereYuk}).

Now, the potential $V_C(x)$ is the one that would have been created by the missing full cylinder due to the presence of the tunnel along the diameter of the sphere. Therefore, the effective gravitational potential felt by the quantum particle at a given distance $x$ from the center of the sphere as it moves inside the cylindrical tunnel is,
\begin{align}\label{VeffYuk}
    V_{\rm eff}(x)&=V_S(x)-V_C(x)\nonumber\\
&=-4\pi G\rho\Bigg[R^2+\frac{x^2}{3}+\alpha\lambda^2e^{-\frac{R}{\lambda}}\cosh\left(\frac{x}{\lambda}\right)-\frac{\alpha\lambda^2}{x}(R+\lambda)e^{-\frac{R}{\lambda}}\sinh\left(\frac{x}{\lambda}\right)\nonumber\\
&\quad-\frac{R-x}{4}\sqrt{(R-x)^2+a^2}-\frac{R+x}{4}\sqrt{(R+x)^2+a^2}\nonumber\\
&\quad-\frac{a^2}{4}\ln\frac{R-x+\sqrt{(R-x)^2+a^2}}{a}-\frac{a^2}{4}\ln\frac{R+x+\sqrt{(R+x)^2+a^2}}{a}\Bigg].
\end{align}

We thus see that, in contrast to the usual exponentially suppressed Yukawa correction to the potential felt by a particle outside a massive sphere, a particle traveling through the tunnel across the sphere feels a Yukawa correction that also contains terms that are proportional to the exponential $e^{x/\lambda}$, as well as terms inversely proportional to the distance $x$ of the particle from the center of the sphere. As we shall see by computing the phase shift, these crucial terms help amplify the contribution from the Yukawa correction to the potential by giving rise to a contribution that is not exponentially suppressed anymore. As a result, this setup does --- thanks to its peculiar configuration --- enhance the effect on quantum particles of a Yukawa-like correction to the gravitational interaction.

Indeed, going back now to our expression for the phase shift and substituting the effective potential (\ref{VeffYuk}), we find the following phase shift,
\begin{align}\label{PhaseYuk}
    \Delta\phi&=-\frac{m^2\lambda_0}{\pi\hbar^2}\int_0^{R}V_{\rm eff}(x){\rm d}x\nonumber\\
    &=\frac{2Gm^2\rho\lambda_0}{\hbar^2}\Bigg[\frac{20R^3}{9}-\frac{(2R^2-a^2)}{3}\sqrt{4R^2+a^2}-\frac{a^3}{3}-a^2R\ln\frac{2R+\sqrt{4R^2+a^2}}{a}\nonumber\\
    &\quad+2\alpha\lambda^3e^{-\frac{R}{\lambda}}\sinh\left(\frac{R}{\lambda}\right)-2\alpha\lambda^2(R+\lambda)e^{-\frac{R}{\lambda}}{\rm Shi}\left(\frac{R}{\lambda}\right)\Bigg].
\end{align}
Here, ${\rm Shi}(z)$ is the hyperbolic sine integral \cite{FormulasBook}. Note that we were able to apply here the prescription described in the previous section ---which consists in taking twice the integral over $x$ between $0$ and $R$--- because the effective potential (\ref{VeffYuk}) is even in the integration variable $x$. In fact, the value of the integral is then odd in the variable $x$ and, hence, simply doubles when taken between the bounds of integration $-R$ and $R$. In this formula, we have a net separation between the contribution of the Newtonian potential and the contribution of the Yukawa-like  correction. We can now find an estimate for the phase shift for any given value of the range $\lambda$ and of the coefficient $\alpha$. 

For a Yukawa range of the order of $\sim10^{-3}\,{\rm m}$, using cold neutrons for which the diameter of the tunnel is only required to be much larger than their $\sim10^{-15}\,{\rm m}$-Compton wavelength, we may easily extract from Eq.~(\ref{PhaseYuk}) an approximation for the phase shift. Indeed, keeping in mind that $a,\lambda\ll R$, as well as $a\ll\lambda$ \cite{TestsReview2014}, we are led, after expanding ${\rm Shi}(R/\lambda)$, to the following expression,
\begin{align}\label{PhaseYukApproximate}
    \Delta\phi&\approx\frac{4GMm^2\lambda_0}{3\pi\hbar^2}\left(1- \frac{9\alpha\lambda^4}{4R^4}\right).
\end{align}
We have expressed here the final result in terms of the mass $M$ of the sphere. This expression shows that the phase shift nicely splits into an overall factor that depends on the mass of the sphere and terms that solely depend on the Yukawa range and the composition-depend factor $\alpha$. As such, it becomes remarkably possible to test out the WEP by placing on both arms of the interferometer massive spheres of the same radius and mass, but of different material compositions. With such a configuration of the interferometer, any phase shift between the two beams would betray a composition-depend gravitational interaction.

Now, although the result (\ref{PhaseYukApproximate}) does not display the usual exponentially decaying term with the radius of the sphere, $e^{-R/\lambda}$, the fact that the correction is proportional to the fourth power of $\lambda$ still makes the phase shift correction extremely small. For a $1\,$m-radius sphere of platinum, and with cold neutrons of wavelength $\lambda_0\sim10\,\si{\angstrom}$, the Newtonian phase shift is of the order of  $0.01\,{\rm rad}$. For an $\alpha$ of the order of $10^{30}$ and a Yukawa range of the order of $10^{-12}\,{\rm m}$ \cite{TestsReview2014}, the correction to this phase shift induced by the Yukawa interaction is of the order of $10^{-18}$. However, for an $\alpha$ of the order of $10^{-2}$ and a range of order $10^{-3}\,{\rm m}$ \cite{TestsReview2014}, the correction is of the order $10^{-14}$. Yet, for even higher values of $\lambda$ (of the order of $10^3\,{\rm m}$) but small values of $\alpha$ of the order of $\sim10^{-7}$, a case which is still not excluded by experiments \cite{TestsReview2014}, expanding formula (\ref{PhaseYuk}) for small $R/\lambda$, yields a correction proportional to $-2\alpha\lambda R^2e^{-R/\lambda}$. The correction we obtain for the phase shift relative to the one caused by ordinary Newtonian gravity in this case becomes then as high as $2\times10^{-4}$.

\subsection{With a power-law deviation}\label{sec:IIIB}
To compute the phase shift between the two beams that would result from an ISL deviation of the form (\ref{PL}), we proceed in a way similar to what we just did with the Yukawa-like correction (\ref{Yukawa}). We first compute the gravitational potential felt by the particle at any distance $x$ from the center of the sphere as it travels through the tunnel. 

First, we can easily see that the correction term in formula (\ref{PL}) gives the gravitational potential at a distance $x$ along the axis of symmetry of a ring of uniform linear mass density $\mu$ and radius $y$ in the form $-2\pi G\mu r_0^{n}\,y/(x^2+y^2)^{\frac{n+1}{2}}$. By integrating this expression over the concentric rings that make up a uniform disk, we easily deduce also the correction to the gravitational potential at a distance $x$ along the axis of symmetry of a massive disk of radius $y$ and of uniform surface mass density $\sigma$. By distinguishing the cases $n=1$ and $n\neq1$, we find the expression $-2\pi G \sigma r_0\ln(\sqrt{x^2+y^2}/x)$ for the former and the expression $-2\pi G\sigma r_0^n[(x^2+y^2)^{(1-n)/2}-x^{1-n}]/(1-n)$ for the latter. 

Using these expressions coming from the uniform massive disk, we can easily compute the gravitational potential $V_C(x)$ at any distance $x$ from the middle point (along the axis of symmetry) of a cylinder of radius $a$, of uniform density $\rho$ and of length $2R$. Using the same strategy adopted for the sphere in Section \ref{sec:IIIA}, these potentials of the uniform disk also allow us to compute the potential $V_S(x)$ at any distance $x$ from the center of a sphere of mass density $\rho$ and radius $R$. Because of the presence of $1-n$ in the denominator of the potential due to a disk, however, we should distinguish between all six different cases of $n=1$ to $n=6$. The corrections $V_{C_{\rm corr}}(x)$, $V_{S_{\rm corr}}(x)$ and $V_{\rm eff_{\rm corr}}(x)$ to the potential of a full cylinder, of a full sphere and of the effective potential inside the tunnel, respectively, emerging from these different cases are calculated in Eqs.~(\ref{CylinderPLn1})-(\ref{VeffPLn6}) of \ref{B}. As can be seen with those results, only the cases $n=1,2,3$ yield converging potentials for all values of $x$. Those are given in Eqs.~(\ref{VeffPLn1}), (\ref{VeffPLn2}) and (\ref{VeffPLn3}). The other three cases of $n$, given in Eqs.~(\ref{VeffPLn4}), (\ref{VeffPLn5}) and (\ref{VeffPLn6}) are all diverging at $x=R$. The divergences for those cases come from the rapidly decreasing potential with distance. The divergence for the cases $n=2,3$ is cured by the presence of the empty tunnel inside of the sphere. In fact, although the expressions (\ref{CylinderPLn2}) and (\ref{SpherePLn2}) for the case $n=2$ contain the diverging integrals $\int_0^{R\pm x}{\rm d}s/s$, the resulting effective potential inside the sphere with a tunnel is finite. Similarly, the potential that results from combining the expressions (\ref{CylinderPLn3}) and (\ref{SpherePLn3}) is finite although both contain the terms $\int_0^{R\pm x}{\rm d}s/s^2$. In contrast, for the more rapidly decreasing potentials of the cases $n=4,5,6$, it does not help much to have an empty tunnel inside the sphere to remedy the short-distance divergences as the latter would have considerably accumulated by the time the particle reaches the outside surface of the sphere. As such, only the cases $n=1,2,3$ will be dealt with here as they lead to finite phase shifts as well. We shall come back on this divergence for the cases $n=4,5,6$ in Section\,\ref{Summary}.

Using the corrections to the effective potentials found in \ref{B}, we compute the following corrections to the phase shifts caused by the non-Newtonian part of the potential for the cases $n=1,2,3$\footnote{Note that the same remark made above also applies here. That is, we are able to apply the prescription that consists in taking twice the integral over $x$ between $0$ and $R$ because all the effective power-law potentials (\ref{VeffPLn1}), (\ref{VeffPLn2}) and (\ref{VeffPLn3}) are even in the integration variable $x$.}.
\begin{align}\label{PhasePL1}
    \Delta\phi_{\rm corr}^{n=1}&=\frac{Gm^2\rho\lambda_0r_0}{\hbar^2}\Bigg[R^2\left(1+\frac{\pi^2}{4}\right)-4Ra\tan^{-1}\left(\frac{2R}{a}\right)-2R^2\ln\frac{4R^2+a^2}{4R^2}+\frac{a^2}{2}\ln\frac{4R^2+a^2}{a^2}\Bigg]\nonumber\\
    &\approx\frac{3GMm^2\lambda_0r_0}{4\pi\hbar^2R}\left(1+\frac{\pi^2}{4}\right).
\end{align}
In the last line, we have expanded the final result and kept the leading-order terms inside the square brackets of the first line. Thus, in contrast to the Yukawa-like correction (\ref{Yukawa}), the power-law correction (\ref{PL}) yields a phase shift that is linear in the ratio $r_0/R$ and is independent of the radius $a$ of the tunnel. 

Similarly, we find for the case $n=2$,  
\begin{align}\label{PhasePL2}
    \Delta\phi_{\rm corr}^{n=2}&=-\frac{Gm^2\rho\lambda_0r_0^2}{\hbar^2}\Bigg[4R+2\sqrt{4R^2+a^2}-2a-4R\ln{\frac{2R+\sqrt{4R^2+a^2}}{a}}\Bigg]\nonumber\\
    &\approx-\frac{3GMm^2\lambda_0r_0^2}{\pi\hbar^2R^2}\left(2-\ln\frac{4R}{a}\right).
\end{align}
In the last line, we have again expanded the final result and kept the leading-order terms inside the square brackets of the first line. In contrast to the case $n=1$, this correction to the phase shift depends on the radius $a$ of the tunnel through a logarithm that involves the ratio of $R$ over $a$. The correction then logarithmically increases with the increase of the relative difference between the radius of the sphere and that of the tunnel inside it.

Finally, for the case $n=3$, we get
\begin{align}\label{PhasePL3}
    \Delta\phi_{\rm corr}^{n=3}&=-\frac{Gm^2\rho\lambda_0r_0^3}{\hbar^2}\Bigg[\frac{\pi^2}{4}-\frac{2R}{a}\tan^{-1}\left(\frac{2R}{a}\right)+\frac{1}{2}\ln\left(1+\frac{4R^2}{a^2}\right)\Bigg]\nonumber\\
    &\approx\frac{3GMm^2\lambda_0r_0^3}{4\hbar^2aR^2}.
\end{align}
This correction is inversely proportional to the radius of the tunnel and, hence, becomes important for smaller radii of the latter. Given the Compton wavelength limit on the smallness of the radius of the tunnel, heavier atoms and molecules, for which $\lambda_C$ is smaller, are best suited for testing this case of $n=3$.

With the $1\,{\rm m}$-radius sphere and cold neutrons, the correction to the phase shift for $n=1$ already reaches the order of $10^{-14}$ for an $r_0$ as small as $10^{-2}\,{\rm pm}$ \cite{TestsReview2014}, but becomes as high as $10^{-3}$ for an $r_0$ of the order of a millimeter. On the other hand, for an order of magnitude of a millimeter for $r_0$, the correction for the case $n=2$ reaches the order of $10^{-6}$, whereas for the case $n=3$ it reaches an order of magnitude which is only $10^{-9}$ for a diameter of the tunnel as small as a nanometer. Therefore, the limitations of the setup increase with the exponent in the power-law deviation formula.

\section{Phase shift due to gravitomagnetism}\label{sec:IV}
Actually, with the same setup as above it is also possible to probe an eventual gravitomagnetism interaction of a rotating sphere with the quantum spin of particles. Indeed, making the massive sphere of mass $M$ rotate counterclockwise around the positive $x$-axis with an angular frequency $\omega$, the sphere creates a gravitomagnetic field $B_G$ parallel to the $x$-axis and oriented in the positive $x$-direction as well. The gravitomagnetic field outside a sphere of uniform density at any distance $r$ from its center is given by ${\bf B}_G^{\rm Out}=2G({\bf L}r^2-3{\bf r}\,{\bf L}\cdot {\bf r})/c^2r^5$ \cite{CiufoliniWheeler}. Here, $\bf L$ is the angular momentum of the rotating sphere, of magnitude $L=2MR^2\omega/5$, and $c$ is the speed of light. 

On the other hand, the magnitude of the gravitomagnetic field inside a spherical shell of exterior radius $R$ is $B_G^{\rm In}=4GM_{\rm Shell}\omega/(3c^2R)$ at any distance from the center of the shell and it has the same direction as ${\bf B}_G^{\rm Out}$. This expression is used as an approximate value for the gravitomagnetism inside a thin shell \cite{CiufoliniWheeler}. We chose this expression here for want of a better and more rigorous formula for the unknown full field in the interior of the successive continuous shells. Nevertheless, being interested here in the possibility of putting into evidence the gravitomagnetic interaction itself, the precise multiplicative factors, which would result from an exact formula of the interior field $B_G^{\rm In}$, would not change our final qualitative conclusions. Furthermore, it is specifically one of the purposes of our present setup to test gravitomagnetism and the precise form the latter should have inside spherical shells\footnote{Note that if one assumes the analogy between magnetism and gravitomagnetism holds even inside the shell and simply replaces the magnetic moment $\mu_B$ by the angular momentum $L_{\rm Shell}$, then the interior gravitomagnetic field would, as for real magnetism, simply be ${\bf B}_G^{\rm In}=2G{\bf L}_{\rm Shell}/c^2R^3$. Therefore, since for a shell of uniform density $\rho$ and of exterior and interior radii $R$ and $x$, respectively, the moment of inertia is $I_{\rm Shell}=\frac{8\pi}{15}\rho(R^5-x^5)$, one would find, $B_G^{\rm In}(x)=\frac{16\pi G\rho\omega}{15c^2R^3}\left(R^5-x^5\right)$.}. 

We therefore would like to stress and emphasize here the fact that our approach in this section is going to be totally heuristic and that it serves the only goal of offering concrete approximate orders of magnitude to be expected from such a setup aimed at testing the unknown gravitomagnetism phenomenon inside rotating bodies. In fact, although our treatment here is fully heuristic ---given that no viable solution to gravitomagnetism inside matter is available yet\footnote{See Ref.\,\cite{EinsteinInMatter2020} for a very recent work on dealing with Einstein equations inside matter and Ref.\,\cite{EinsteinInMatter1995} for a brief review on the latter topic.}--- the formulas we extract from our present analysis allow us to pinpoint the difference brought, and the advantage offered, by having particles travel through a channel over conventional setups that rely instead on the effect of the external gravitomagnetic field of rotating bodies when it comes to testing gravitomagnetism. Our subsequent formulas displayed below should thus be taken with a grain of salt as they are not based on any known rigorous expression of the eventual gravitomagnetic field inside the solid rotating sphere. Our approach here will indeed consist in desperately resorting to the superposition principle suggested only by the weakness of the approximate gravitomagnetic field expression inside a spherical thin shell we displayed above and which we borrowed from Ref.\,\cite{CiufoliniWheeler}\footnote{The weakness of the gravitomagnetic field has actually been exploited for other purposes in literature. In the recent work in Ref.\,\cite{QCOblate}, the weak-field limit has been exploited to study quantum corrections to the time delay caused by a spinning source and in the more recent work in Ref.\,\cite{NonlocalGM} it has been used to estimate the correction to the external gravitomagnetism that would arise from nonlocal gravity.}. Our formulas do, nevertheless, serve well the purpose of pointing towards the best possible way of efficiently exploiting quantum particles when investigating gravitomagnetism, especial the one that could be created inside rotating bodies.

Being interested here in neutral particles with spins aligned along the positive $x$-axis and traveling along the axis of the sphere, we only need the magnitudes of such gravitomagnetic fields at the north pole of both the sphere and the shell. At that point, we have ${\bf r}$ parallel to $\bf L$ so that the magnitude of the exterior field of a sphere of radius $r$ is $B_G^{\rm Out}=8GM\omega/(5c^2r)$.
Now, as the particle is traveling through the channel, the gravitomagnetic field it feels changes with its position $x$ from the center. In fact, as the particle moves forward, a $B_G^{\rm Out}$ at position $x$ is created by the inner sphere of radius $x$ and mass $4\pi\rho x^3/3$ in front of the particle. Another field $B_G^{\rm In}$ is created by a shell, of exterior radius $R$ and interior radius $x$, surrounding the particle at that specific position $x$. The mass of a shell of uniform density $\rho$ and of exterior and interior radii $R$ and $x$, respectively, is given by $M_{\rm Shell}=\frac{4\pi}{3}\rho(R^3-x^3)$. Therefore, the total gravitomagnetic field felt by the particle at a position $x$ from the center inside the tunnel would be,
\begin{align}\label{BG}
    B_G(x)&=\frac{32\pi G\rho\omega x^2}{15c^2}+\frac{16\pi G\rho\omega}{9c^2R}\left(R^3-x^3\right).
\end{align}

Thus, for a beam of polarized neutrons such that the spins of those going through the channel are perfectly aligned with the positive $x$-direction, we have the gravitomagnetism (GM) interaction Hamiltonian $H_{GM}(x)=sB_G(x)$, where $s=\hbar/2$ is the intrinsic spin of the polarized particles. Now, the phase difference induced by any interaction Hamiltonian between times $t_i$ and $t_f$ is given by $\Delta\phi=\hbar^{-1}\int_{t_i}^{t_f}H(x){\rm d}t$ \cite{COWReview}. Thus, in addition to the phase differences found above for a nonrotating sphere, the rotation of the sphere would induce an extra phase difference $\Delta\phi_{GM}$ between the two beams given by,
\begin{align}\label{PhaseGM}
    \Delta\phi_{GM}&=\frac{16\pi G\rho\omega}{3c^2}\int_{t_i}^{t_f}\left(\frac{x^2}{5}+\frac{R^2}{6}-\frac{x^3}{6R}\right){\rm d}t\nonumber\\
    &=\frac{32\pi G\rho\,m\lambda_0\omega }{3h c^2}\int_{0}^{R}\left(\frac{x^2}{5}+\frac{R^2}{6}-\frac{x^3}{6R}\right){\rm d}x\nonumber\\
    &=\frac{23GM\,m\lambda_0\omega }{15h c^2}.
\end{align}
In the second line we have traded ${\rm d}t$ for ${\rm d}x$ by introducing the neutrons' velocity $v_0=h/m\lambda_0$ considered, within our approximations, to be constant all along the tunnel. We have multiplied by a factor of 2 as our integration variable $x$ goes from $0$ to $R$ whereas the particle feels a symmetric potential on both sides of the center of the sphere along its trip through the tunnel. In other words, having taken here twice the integral over $x$ between $0$ and $R$ is justified not because of the parity of the integrated function, as was the case for the gravitational potentials of the previous section, but because $x$ represents here the distance of the particle from the origin. We have expressed the final result in terms of the mass $M=4\pi\rho R^3/3$ of the sphere instead of its radius and mass density\footnote{Had we relied on a pure analogy between magnetism and gravotomagnetism, we would have found the phase shift $\frac{32\pi G\rho\,m\lambda_0\omega }{15h c^2}\int_{0}^{R}\left(x^2+\frac{R^2}{2}-\frac{x^5}{2R^3}\right){\rm d}x= \frac{8GMm\lambda_0\omega}{5h c^2}$. Thus, we clearly see indeed that only the numerical factors would differ in the final result.}. The benefit of having the particles travel through the channel is obviously to accumulate the induced phase differences, making the total phase difference scale like $\rho R
^3$ instead of $\rho R^2$ as it would be the case when relying on the purely external gravitomagnetic field. Furthermore, we have not included here the phase shift due to the extra distance the particle travels outside the sphere, first, from the reflector on the left all the way to the sphere, and then from the sphere all the way to the beam splitter on the right. If the reflector and the beam splitter are each at a distance $L$ from the center of the sphere, then the extra phase shift brought by the extra path is given by, 
\begin{equation}\label{PhaseGMExtra}
    \Delta\phi_{GM}^{\rm Out}=\frac{4 GM\omega}{5c^2}\int_{t_i}^{t_f}\frac{{\rm d}t}{x}=\frac{8 GM\,m\lambda_0\omega }{5h c^2}\int_{R}^{L}\frac{{\rm d}x}{x}=\frac{8GM\,m\lambda_0\omega }{5h c^2}\ln\frac{L}{R}.
\end{equation}
This extra phase shift is clearly very small whenever the reflector and the beam splitter are not very far from the surface of the sphere.

These formulas for the phase shift show that, in addition of requiring a very massive and a very fast rotating sphere, the longer is the wavelength $\lambda_0$, i.e., the colder are the particles, the bigger the measured shift will be. With a $1\,{\rm m}$-radius sphere of platinum, rotating at the rate of $300$ revolutions per minute, and cold neutrons of wavelength $\lambda_0\sim10\si{\angstrom}$, the induced phase shift due to gravitomagnetism would be of the order of $10^{-19}\,$rad. Increasing the mass of the test particles, as well as the mass and the angular speed of the sphere would, of course, increase the resulting phase shift, and only mechanical limitations could hinder improvements on this latter front due to the minute phase shifts involved. It is, nevertheless, clearly a daunting experimental and engineering challenge to achieve such a rotation rate with such a heavy massive sphere through which cold quantum particles with aligned spins must travel.


\section{The gravitationally induced harmonic oscillator}\label{sec:V}
The possibilities offered by our setup depicted in Fig.~\ref{COWBall} do not end with the interference experiments presented in the previous two sections. In fact, looking at the effective potential in Eq.~(\ref{VeffYuk}), we see that when neglecting the Yukawa-like correction, and for a very small radius $a$ of the tunnel, the gravitational potential simplifies considerably and yields, up to a constant term,
\begin{equation}
    V_{\rm eff}(x)=\frac{2\pi G\rho}{3} x^2.
\end{equation}
This is nothing but the potential of a simple harmonic oscillator with one degree of freedom along the $x$-axis. No charge is required to be carried by the test particles in this case. Therefore, any neutral quantum object of a size ranging from the that of a neutron, to that of atoms and molecules, and all the way to the size of objects exhibiting macroscopic quantum states like superfluids\footnote{An investigation of the effect of such a gravitationally induced harmonic oscillator on the super-electrons inside a superconductor is under investigation by the present authors.}, could very well serve our purpose. For simplicity, though, we just continue to assume here that one still uses cold neutrons as test particles. The energy of such neutrons becomes then quantized inside the tunnel and is given by that of a simple harmonic oscillator:
\begin{equation}\label{SHM}
    \mathcal{E}=\hbar\omega\left(n+\frac{1}{2}\right),
\end{equation}
with $n$ a non-negative integer and the fundamental angular frequency is given by $\omega=(\frac{4\pi}{3} G\rho)^{1/2}$. These quantized energies do not thus depend on the size of the sphere but only on the density of the latter. This makes the experimental setup more flexible. For a platinum sphere of density $2.145\times10^{4}\,{\rm kg/m^3}$, the predicted difference between two consecutive energy levels is of the order of $10^{-18}\,{\rm eV}$.  

Now, for very small displacements $x\ll\lambda$ away from the center inside the tunnel, we may extract from Eq.~(\ref{VeffYuk}) the following approximation for the gravitational potential of the neutrons due to the Yukawa-like deviation:
\begin{equation}\label{SHMYuk}
    V_{\rm eff}(x)=\frac{2\pi G\rho}{3}x^2 \left[1-2\alpha\left(1-\frac{R}{2\lambda}\right)e^{-R/\lambda}\right].
\end{equation}
This correction to the potential of the oscillator yields the following modification to the fundamental angular frequency of the particles, $\omega\approx(\frac{4\pi}{3} G\rho)^{1/2}[1-\alpha(1-\frac{R}{2\lambda})e^{-\frac{R}{\lambda}}]$. This correction depends on the size of the sphere. Moreover, the correction is exponentially suppressed for ranges of the order of the micrometer. Only for ranges of $\lambda$ of the order of $R$ and beyond do we get a correction of the order of the factor $\alpha$.

Still, for very small $x\ll R$ again, and using Eq.~(\ref{VeffPLn1}), we get, up to a constant, the following expression for the gravitational potential of the harmonic oscillator due to the power-law (\ref{PL}) with $n=1$:   
\begin{equation}\label{SHMPLn1}
    V_{\rm eff}^{n=1}(x)=\frac{2\pi G\rho}{3}x^2 \left(1+\frac{2r_0}{R}\right).
\end{equation}
We see that the correction is simply of the order of $r_0$ for a unit-radius sphere. For smaller radii of the sphere, the correction is enhanced as it is inversely proportional to the radius $R$. Similarly, for the case $n=2$ we find, the following expression, after expanding (\ref{VeffPLn2}) at the first order in $x$ and discarding constant terms:
\begin{equation}\label{SHMPLn2}
    V_{\rm eff}^{n=2}(x)=\frac{2\pi G\rho}{3}x^2 \left(1+\frac{3r_0^2}{R^2}\right).
\end{equation}
This expression displays a correction that is still independent of the radius of the tunnel and a proportionality to the square of the ratio $r_0/R$. Like with the case of $n=1$, the smaller is the sphere used in the experiment, the more enhanced will be the correction. Finally, for the case $n=3$, a similar approximation for small $x$ as the one done for the previous two cases of $n$ leads to the following correction:
\begin{equation}\label{SHMPLn3}
    V_{\rm eff}^{n=3}(x)=\frac{2\pi G\rho}{3}x^2 \left(1+\frac{r_0^3}{R^3}\right).
\end{equation}
Thus, in all three cases of the power-law deviation (\ref{PL}), the order of magnitude of the correction to the gravitational potential of the induced simple harmonic oscillator is conditioned by the order of magnitude of the ratio $r_0/R$.
\section{A variant of the setup}\label{sec:VI}
In the previous sections, we showed how the cylindrical tunnel drilled inside the massive sphere helps induce quantum interference between the arms of the interferometer. The advantage of the tunnel is that it offers a pathway to the particles inside the sphere, making the particles interact with every layer of the latter. The technological challenge behind such a setup consists, of course, in making as many neutrons as possible go through such a small-diameter hole and the difficulty in detecting the small induced phase shifts. In this section, we are going to describe a variant of the setup that is based on the same principle as the previous one but in which the resolution could be improved and the limitations corresponding to the focusing of the neutrons could be overcome.

In fact, a variant of such a setup consists simply in replacing the cylindrical tunnel inside the sphere by an empty disk of thickness $2a$. That is, the sphere becomes then split into two hemispheres of the same radius and mass, between which the particles could freely move. As we shall see now, this configuration allows us to easily find an estimate for the Yukawa-like correction without putting such a high constraint on the separation distance $2a$ between the two hemispheres. In addition, allowing larger values of $a$ removes the constraints in the focusing of the neutron beam.    

With a Yukawa-like correction (\ref{Yukawa}), the gravitational potential $V_{H\rm eff}(x)$ between the two hemispheres at a distance $x$ from the center is given by $V_S(x)-V_D^N(x)-V_D^Y(x)$, where the potentials $V_S(x)$, $V_D^N(x)$ and $V_D^Y(x)$ are given by Eqs.~(\ref{SphereYuk}), (\ref{DiskNew}) and (\ref{DiskYuk}), respectively. Then, using Eqs.~(\ref{DiskNewIntegrated}) and (\ref{DiskYukIntegrated}), we find the induced phase shift on the neutrons travelling between the two hemispheres as follows, 
\begin{align}\label{PhaseDiskYuk}
    \Delta\phi_H&=-\frac{m^2\lambda_0}{\pi\hbar^2}\int_0^{R}V_{H\rm eff}(x){\rm d}x\nonumber\\
    &\approx\frac{2Gm^2\rho\lambda_0}{\hbar^2}\left[\frac{8R^3}{9}-aR\left(RI_D^N-a+2\lambda\alpha e^{-a/\lambda}\right)\right].
\end{align}
As indicated in \ref{C}, what is contained inside the factor $I_D^N$ depends on the degree of precision one would want to achieve which, in turn, depends on the order of magnitude of $a$. However, for an $\alpha$ of the order $10^{-2}$ and a Yukawa range $\lambda$ of the order of $10^{-3}\,{\rm m}$, the phase shift correction due to the Yukawa-like contribution is already of the order $\sim10^{-5}$ for 1\,m-radius hemispheres and for a separation distance $2a$ between them of the order of a millimeter. For this reason, only the first-order term in $\lambda$ is kept here. For higher values of $\alpha$ \cite{TestsReview2014} the correction becomes, of course, enhanced. Comparing formulae (\ref{PhaseYukApproximate}) and (\ref{PhaseDiskYuk}), we clearly see the advantage of having the particles go in between the two hemispheres. It helps avoid the fourth power of $\lambda$ and replaces it instead by the exponential $e^{-a/\lambda}$, which calls, of course, for a small separation $2a$ between the two hemispheres. Yet, there is still the advantage of having a tunnel inside a massive sphere instead of just two separated hemispheres as the former provides a guide for the particles as well.


\section{Summary and discussion}\label{Summary}
Our simple setup could constitute a new tool for contributing to the various tests on the interplay between gravity and quantum mechanics. The basic principle of the setup is similar to that of a COW experiment in that it relies on the quantum interference between the two beams of a Mach-Zehnder interferometer. The difference with a COW experiment lies in the fact that now the test particles interact in a different way with the mass source. The tunnel through the mass source (or the regions between the two massive hemispheres) allows test particles to interact fully with {\it every layer} of the mass. The resulting final phase shift between the two beams is thus due to an accumulation of infinitesimal phase shifts acquired by the particle along its journey inside the tunnel (or between the two hemispehres). This accumulation effect helps to improve the sensitivity of existing experiments aimed at testing deviations from the ISL using tabletop massive sources of gravity and other means \cite{Constraints2004,Constraints2005,YukawaIJMP2011,f(R)Yukawa,Saha2013,NewConstraints}. 

Now, one might argue that since any deviation from the ISL would automatically have a short interaction range, one could then just replace the heavy sphere by a long thin tube through which the particle would travel as the latter would only interact with the immediate surrounding material from which the tube is made. This could indeed be done in practice, but using a long thin tube would not have the same ``using up'' effect as the one provided by a full solid sphere with a hole. To see this, one can just use Eq.\,(\ref{CylinderYuk}) which gives the potential inside a full cylinder. To find the potential inside a thin tube of thickness, say $b-a$, one has only to subtract the potential given by Eq.\,(\ref{CylinderYuk}) with a diameter $2a$ of the cylinder from the potential given by Eq.\,(\ref{CylinderYuk}) with a diameter $2b$ of the cylinder. The resulting Yukawa deviation is clearly not the same and is less important than what is found in Eq.\,(\ref{VeffYuk}) for a sphere with a hole. This can physically be understood as being due to an accumulation of the (albeit small) contributions from the successive concentric spheres through which the quantum particle goes during its journey through the hole. This is also valid for the power-law deviation from the Newtonian potential. Thus, the capacity of using up the whole source mass comes both from (i) the contribution of the many concentric spheres along the journey of the particle and (ii) the fact that the phase shifts are accumulated within the prolonged duration it takes the particle to complete its journey.

It was possible to examine here both a Yukawa-like deviation form the ISL and a power-law deviation. For the latter, only the cases $n=1,2,3$ yield a finite potential felt by the particle as it travels through the tunnel inside the sphere. Now, the fact that we obtained divergent effective gravitational potentials inside the tunnel for the cases $n=4,5,6$ is in itself a very interesting physical result ---regardless of its use for interferometry purposes--- and is worth discussing here. These divergences are due to the much more rapidly increasing potentials with the decreasing distance for the cases $n=4,5,6$. When taken at face value, this fact alone already constitutes a solid physical argument against the validity of such power-law deviations from the ISL at very short distances. The only cure for such a pathology in this case is, indeed, to have such a power-law deviation from the ISL be itself modified and corrected at even much shorter distances. The natural conclusion then would be that these power-law models themselves, if valid at certain short distances, would certainly be accompanied by new deviations at even shorter distances as they cannot be valid at all scales. In any case, our proposed setup could allow at least to detect possible interferometric effects arising from power-law deviations with $n=1,2,3$ and, at the same level of importance, to measure the gravitational constant $G$ to unprecedented precision and to study its behavior at very short distances.

The only requirement one needs to respect with such a setup is to make the diameter of the tunnel (or the separation of the two hemispheres) quite larger than the Compton wavelength of the particles in order to avoid unwanted diffraction of the particles when they come out of the sphere. For cold neutrons at the lowest temperatures of the order of $10^{-3}\,{\rm K}$ presently achievable, the diameter of the tunnel is allowed to be very small indeed. 

We would like to stress here that our setup is designed to work with any quantum object as a test particle since the only important property of such objects is their matter waves and the interference they exhibit. As such, atom interferometers, or even flowing superfluids, could just as well do the job. Presently available small-scale neutron interferometers are thus neither necessary nor the best for the task. In fact, given the present precision and limitations of such devices (sizes of a few centimeters and phase shift measurements precision of the order of $10^{-4}\,$rad), the goals addressed here seem to be out of reach when based on such presently available designs of this kind of interferometers. The presently widely used designs for neutron interferometery are indeed based on Bragg diffraction obtained from silicon-crystal blades carved from a monolithic base made of a single silicon ingot of the order of a few centimeters (see, e.g., Ref.\,\cite{Overview}). Similar size limitations of the order of a centimeter are also encountered in the three phase-grating moir\'{e} neutron interferometers supposed to cover larger areas \cite{ThreePhaseGrating}. This far-field neutron interferometry technique allows indeed to fully use intense neutron sources for precision measurements \cite{FarField}. In addition, this technique would also serve well our needs as it overcomes the alignment and stability issues, as well as the fabrication challenges associated with the more conventional perfect-crystal neutron interferometers \cite{FarField}. However, as our setup requires a large mass source, even this kind of neutron interferometer could not be used as it is without modification. Nevertheless, the technology employed there does actually allow to reach distances of the order of a few meters between the slit receiving the incident beam and the imaging camera recording the fringes pattern due to the moir\'{e} effect \cite{FarField}. Therefore, while such interferometers themselves cannot be used as they are for our purposes, the technology on which they rely can be adopted to our setup to provide the needed spacing, intensity, stability and alignment of the neutron beams.

Furthermore, the same setup could also help, as we saw, to gravitationally induce a quantum harmonic oscillator. The Yukawa-like correction and the power-law deviation bring distinct modifications to the quantized energy of the harmonic oscillator. The contribution of the former is exponentially suppressed with the size of the sphere but the contribution of the latter depends on the ratio $r_0/R$ with a power equal to the power $n$ in the formula (\ref{PL}). The cases $n=4,5,6$ do not yield a finite potential for the particle inside the tunnel. As emphasised above, the use of neutrons to exploit such a gravitationally induced harmonic oscillator is not necessary. In fact, given the difficulty in handling a neutron cloud inside such a narrow tunnel, other quantum objects like atoms and even superfluids could be used instead.

In addition, our setup offers the possibility of testing gravitomagnetism in a novel way by measuring the {\it accumulated} phase shift a quantum particle acquires due to each layer of the rotating source as it travels through the tunnel. Moreover, as the particle moves along the diameter of the sphere, around which the latter is rotating, the coupling between the gravitomagnetic field and the intrinsic spin of the polarized particles of the beam is at its maximum. This offers the best configuration ever for testing gravitomagnetism. Unfortunately, as we saw, the coupling between gravitomagnetism and intrinsic spin of particles is still very small. However, our setup offers the possibility of improving the sensitivity of tabletop experiments aimed at testing gravitomagnetism by using in the future polarized atoms and molecules by relying on their total intrinsic spin and orbital angular spin \cite{AtomsMolecules2017,WithAtoms1,WithAtoms2}.

It is well-known that it is often the case that minute phase shifts are involved in any quantum interference experiment designed to take into account gravitational effects \cite{Nandi,Werner,Okawara1,Okawara2,Okawara3}. In this regard, our setup is no exception. Therefore, our setup heavily relies on future improvements in neutron interferometry technology. Achieving decoherence-free environments and high levels of noise suppression, based, for example, on quantum-error-correcting codes \cite{Pushin}, remains indeed very critical. In particular, a neutron interferometry that would include quantum error corrections to protect our experimental setup against mechanical vibrations \cite{Pushin}, that accompany Mach-Zehnder configurations, is very much needed, especially when the mass source is required to be spinning as is the case when testing gravitomagnetism. Increasing phase shifts measuring precision entails the possibility of using smaller-scale source masses which, in turn, lead to a better control of noise levels and an easier achievement of coherence-free environments.

Now, it should be noted that although our setup relies here solely on the interference collected at the end of the second beam splitter like in a COW experiment, it is not excluded that the same setup be combined with other more precise techniques for measuring minute energy differences such as the use of Ramsey interferometry instead \cite{WithRamsey}. A future work relying on this setup, but combining modern techniques for measuring small phase shifts, will be devoted to reach such a goal. 

\section*{Acknowledgments}
The authors are grateful to the anonymous referee for the very pertinent and helpful comments. This work is supported by the Natural Sciences and Engineering Research
Council of Canada (NSERC) Discovery Grant (RGPIN-2017-05388).
\appendix
\section{The gravitational potential inside the tunnel based on formula (\ref{Yukawa})}\label{A}
In this appendix we are going to compute the required gravitational potentials $V_C(x)$ and $V_S(x)$ needed in subsection \ref{sec:IIIA}. These are due, respectively, to the a full uniform cylinder of radius $a$ and length $2R$ and a full uniform sphere of radius $R$; both having the same uniform density $\rho$. As explained in subsection \ref{sec:IIIA}, we shall use the gravitational potential $-2\pi G\sigma[\sqrt{x^2+y^2}-x+\alpha\lambda(e^{-x/\lambda}- e^{-\sqrt{x^2+y^2}/\lambda})]$ at a distance $x$ along the axis of symmetry of a uniform disk of radius $y$ and surface mass density $\sigma$. By integrating this potential over the region $s\in[0,R-x]$, which lies to the left of the point $x$, and then over the region $s\in[0,R+x]$ which lies to the right of the point $x$, we find, after setting for convenience, $f(s)=\sqrt{s^2+a^2}-s$ and $g(s)=\exp(-s/\lambda)-\exp(-\sqrt{s^2+a^2}/\lambda)$,
\begin{align}\label{CylinderYuk}
V_C(x)&=-2\pi G\rho \left(\int_0^{R-x}+\int_0^{R+x}\right)f(s)\,{\rm d}s-2\pi G\rho\alpha\lambda\left(\int_0^{R-x}+\int_0^{R+x}\right)g(s)\,{\rm d}s\nonumber\\
&=-\pi G\rho\Bigg[(R-x)\sqrt{(R-x)^2+a^2}-(R-x)^2+(R+x)\sqrt{(R+x)^2+a^2}\nonumber\\
&-(R+x)^2+a^2\ln\frac{R-x+\sqrt{(R-x)^2+a^2}}{a}+a^2\ln\frac{R+x+\sqrt{(R+x)^2+a^2}}{a}\nonumber\\
&+2\alpha\lambda\left(2\lambda\!-\lambda e^{-\frac{R-x}{\lambda}}\!-\lambda e^{-\frac{R+x}{\lambda}}\!-I[x,a,R,\lambda]\right)\!\!\Bigg].
\end{align}
The four first lines in this result are solely due to the Newtonian potential, whereas the very last line represents the correction due to the Yukawa-like deviation. 

Here, $I[x,a,R,\lambda]$ arises from the sum of the two integrals in the second line, involving the exponential, $(\int_0^{R- x}+\int_0^{R+ x})e^{-\sqrt{s^2+a^2}/{\lambda}}{\rm d}s$. Such an integral does not admit any analytical expression. It is, nevertheless, possible to estimate its order of magnitude. In fact, with the change of variable $s/a=\sinh z$, the integrals acquire the form $(\int_0^{z_{0-}}+\int_0^{z_{0+}})e^{-\frac{a}{\lambda}\cosh z}a\cosh z\,{\rm d}z$, with the boundaries given by $z_{0\pm}=\sinh^{-1}(\frac{R\pm x}{a})$. These integrals have a similar form as the integral giving rise to the modified Bessel function $K_1(\frac{a}{\lambda})$, which reads $\int_0^{\infty}e^{-\frac{a}{\lambda}\cosh z}\cosh z\,{\rm d}z$ \cite{FormulasBook}. The only difference, then, is in their respective ranges of integration. Given that our integrals stop at the finite boundaries $z_{0\pm}$, and the fact that the factor $e^{-\frac{a}{\lambda}\cosh z}$ is rapidly suppressed for positive values of $z$, it is clear that the value of the sum of our two integrals would not differ much from the value of the modified Bessel function $K_1(\frac{a}{\lambda})$. On the other hand, we know that for large arguments $u$, we have the approximation $K_1(u)\sim\sqrt{\frac{\pi}{2u}}e^{-u}$ \cite{FormulasBook}. Therefore, we deduce that the integral $I[x,a,R,\lambda]$ is of the order of $\sqrt{a\lambda}e^{-\frac{a}{\lambda}}$. By adopting this value, we will be off from the real value, at most, by terms of the form $\sqrt{a\lambda}e^{-\sqrt{(R\pm x)^2+a^2}/\lambda}$. However, for very small values --- compared to a typical Yukawa range $\lambda$ --- of the radius $a$ of the tunnel (which is only much larger than the Compton wavelength of the test particles, $\sim10^{-15}\,{\rm m}$ for neutrons), these terms are much smaller than the rest of the terms in the last line of the result (\ref{CylinderYuk}). For this reason, we may safely discard the term $I[x,a,R,\lambda]$ from the potential (\ref{CylinderYuk}) due to the cylinder.

By making use of the same expression we found above for the gravitational potential of a massive disk at a distance $x$ along its axis, we also easily compute the gravitational potential at a distance $x$ from the center of a sphere of radius $R$ and of uniform density $\rho$. For that purpose, we set for convenience $f(s,y)=\sqrt{s^2+y^2}-s$ and $g(s,y)=\exp(-s/\lambda)-\exp(-\sqrt{s^2+y^2}/\lambda)$. Then, integrating again separately over the left and right regions with respect to the point $x$, we find,
\begin{align}\label{SphereYuk}
V_S(x)&=-2\pi G\rho \left(\int_0^{R-x}\!+\int_0^{R+x}\right)f(s,y)\,{\rm d}s\!-\!2\pi G\rho\alpha\lambda\left(\int_0^{R-x}\!+\int_0^{R+x}\right)g(s,y)\,{\rm d}s\nonumber\\
&=-2\pi G\rho\Bigg[R^2-\frac{x^2}{3}+\alpha\lambda^2\left(2-e^{-\frac{R-x}{\lambda}}-e^{-\frac{R+x}{\lambda}}\right)\nonumber\\
&\quad+\frac{\alpha\lambda^2}{x}\left((R+x+\lambda)e^{-\frac{R+x}{\lambda}}-(R-x+\lambda)e^{-\frac{R-x}{\lambda}}\right)\!\Bigg]
\end{align}
To evaluate the integrals over the volume of the full sphere, we have taken into account the shape of the sphere as follows. For the infinitesimally thin disks that lie on the left region of the point $x$, for which $s\in[0,R-x]$, we used the fact that $y^2=R^2-(s+x)^2$, whereas for the disks that lie on the right region with respect to the point $x$, for which $s\in[0,R+x]$, we used the fact that $y^2=R^2-(s-x)^2$. The first two terms in this result are due to the Newtonian potential, whereas the rest of the terms, which are proportional to $\alpha$, are due to the Yukawa-like deviation. Notice that the potential (\ref{SphereYuk}) is finite everywhere inside the sphere, including the origin. 

The gravitational potential at any distance $x$ from the origin outside a full sphere, based on formula (\ref{Yukawa}), is easily found by adding the Newtonian part $V_S^N(x)$ to the Yukawa part $V_S^Y(x)$. As we saw above, the gravitational potential at a distance $x$ along the axis of symmetry of a uniform disk of radius $y$ and surface mass density $\sigma$ due to the Yukawa-like correction is $-2\pi G\sigma[\alpha\lambda(e^{-x/\lambda}- e^{-\sqrt{x^2+y^2}/\lambda})]$. By integrating this potential over the region $r\in[0,R]$, which lies to the left and to the right of the center of the sphere, and using $y^2=R^2-r^2$, where $r$ is the distance of the disk of radius $y$ from the center of the sphere, we find the following gravitational potential outside the full sphere,
\begin{align}\label{OutsideSphereYuk1}
V_S^{\rm Out}(x)&=V_S^{N({\rm Out})}(x)+V_{S}^{Y({\rm Out})}(x)\nonumber\\
&=-\frac{GM}{x}\!-\!2\pi G\rho\alpha\lambda\int_0^R\Bigg[e^{-(x+r)/\lambda}\!+\!e^{-(x-r)/\lambda}\!-\!e^{-\sqrt{x^2+R^2+2xr}/\lambda}\!-\! e^{-\sqrt{x^2+R^2-2xr}/\lambda}\Bigg]{\rm d}r\nonumber\\
&=-\frac{GM}{x}+4\pi G\rho\alpha \lambda^2e^{-x/\lambda}\left[\frac{\lambda}{x}\sinh\left(\frac{R}{\lambda}\right)-\frac{R}{x}\cosh\left(\frac{R}{\lambda}\right)\right].
\end{align}
Similarly, recalling that the tunnel inside the sphere has the radius $a$ and the length $2R$, we would have the following gravitational potential outside the missing cylinder of radius $a$ and of length $2R$:
\begin{align}\label{OutsideSphereYuk2}
V_C^{\rm Out}(x)&=V_C^{N({\rm Out})}(x)+V_{C}^{Y({\rm Out})}(x)\nonumber\\
&=-2\pi G\rho\int_0^{R}\left[\sqrt{(x+r)^2+a^2}+\sqrt{(x-r)^2+a^2}-2x\right]{\rm d}r\nonumber\\
&\quad-2\pi G\rho\alpha\lambda\int_0^{R}\Bigg[e^{-(x+r)/\lambda}+e^{-(x-r)/\lambda}- e^{-\sqrt{(x+r)^2+a^2}/\lambda}-e^{-\sqrt{(x-r)^2+a^2}/\lambda}\Bigg]{\rm d}r\nonumber\\
&\approx-\pi G\rho\Bigg[(x+R)\sqrt{(x+R)^2+a^2}-(x-R)\sqrt{(x-R)^2+a^2}-4xR\nonumber\\
&\quad+a^2\ln\frac{\sqrt{(x+R)^2+a^2}+x+R}{\sqrt{(x-R)^2+a^2}+x-R}+4\alpha \lambda^2e^{-x/\lambda}\sinh\frac{R}{\lambda}\Bigg].
\end{align}
Here, we have kept only the terms up to the second order in $a$ and the leading exponential terms $e^{-(x\pm r)/\lambda}$. Therefore, the effective gravitational potential outside the sphere with the tunnel drilled inside it is $V_{\rm eff}^{\rm Out}(x)=V_S^{\rm Out}(x)-V_C^{\rm Out}(x)$. This potential yields an extra induced phase shift due to the trip made by the particle outside the sphere from the left reflector to the surface of the sphere and from the surface of the sphere to the right beam splitter, both at a distance $L$ from the center of the sphere. This extra phase shift evaluates to,
\begin{equation}\label{OutsidePhaseYuk}
    \Delta\phi^{\rm Out}=-\frac{m^2\lambda_0}{\pi\hbar^2}\int_R^{L}V_{\rm eff}^{\rm Out}(x){\rm d}x\approx\frac{GMm^2\lambda_0}{\pi\hbar^2}\left[\ln\frac{L}{R}+\frac{3\alpha\lambda^3}{2R^3}e^{(R-L)/\lambda}\left(1-\frac{R}{L}\right)\right].
\end{equation}
Provided that the reflector on the left and the beam splitter on the right are both close to the surface of the sphere, i.e., that $L\sim R$, this extra phase shift will not contribute much to the phase shift (\ref{PhaseYukApproximate}) caused by the interior of the sphere on the particles.

\section{Correction to the gravitational potential inside the tunnel based on formula (\ref{PL})}\label{B}
In this appendix we are going to compute the effective gravitational potential inside the tunnel at any distance $x$ away from the center of the sphere. For that purpose, we follow the same strategy as the one followed in subsection \ref{sec:IIIA} for the Yukawa-like deviation. We first compute the gravitational potential $V_C(x)$ inside a full cylinder of mass density $\rho$, of radius $a$ and length $2R$, at any distance $x$ from the middle point for the cylinder along its axis of symmetry. Then, we compute the potential $V_S(x)$ at any distance $x$ from the center of a full sphere of radius $R$ and of mass density $\rho$. We are going to examine individually each of the different cases $n=1,2,3,4$.
\vspace{-0.5cm}
\subsection{Case: $n=1$.}
We saw in Section~\ref{sec:IIIB} that for the case $n=1$, the correction to the Newtonian gravitational potential due to a disk of radius $y$ and mass density $\sigma$, at a distance $x$ along the axis of symmetry, is  $-2\pi G\sigma r_0\ln(\sqrt{x^2+y^2}/x)$. Let us then introduce, for convenience in order to make the calculations clearer, the function $g(s)=\ln(\sqrt{s^2+a^2}/s)$. Then, by integrating such a function separately over the left and right regions of a point that is at a distance $x$ away from the center of a full cylinder of radius $a$ and of length $2R$, we easily find the following correction $V_{C_{\rm corr}}^{n=1}(x)$ to the Newtonian potential at that point inside the cylinder,
\begin{align}\label{CylinderPLn1}
V_{C_{\rm corr}}^{n=1}(x)&=-2\pi G\rho r_0\left(\int_0^{R-x}+\int_0^{R+x}\right)g(s)\,{\rm d}s\nonumber\\
&=-2\pi G\rho r_0\Bigg[(R-x)\ln\frac{\sqrt{(R-x)^2+a^2}}{R-x}+(R+x)\ln\frac{\sqrt{(R+x)^2+a^2}}{R+x}\nonumber\\
&\quad+a\tan^{-1}\left(\frac{R-x}{a}\right)+a\tan^{-1}\left(\frac{R+x}{a}\right)\Bigg].
\end{align}
In order to compute the correction $V_{S_{\rm corr}}^{n=1}(x)$ to the potential inside a full sphere, we shall use for convenience the function $g(s,y)=\ln(\sqrt{s^2+y^2}/s)$. For the integration, however, we should take again into account the shape of the sphere by using the fact that $y^2=R^2-(s+x)^2$ for the disks on the left of the point $x$ while $y^2=R^2-(s-x)^2$ for the disks on the right of the point $x$. Thus we find,
\begin{align}\label{SpherePLn1}
V_{S_{\rm corr}}^{n=1}(x)&=-2\pi G\rho r_0\left(\int_0^{R-x}+\int_0^{R+x}\right)g(s,y)\,{\rm d}s\nonumber\\
&=-2\pi G\rho r_0\left[R+\frac{R^2-x^2}{2x}\ln\frac{R+x}{R-x}\right].
\end{align}
Combining these corrections to the potential inside the sphere and the cylinder with the Newtonian potential inside each one of the latter, found in \ref{A}, we compute the correction to the effective potential inside the tunnel for case $n=1$ to find,
\begin{align}\label{VeffPLn1}
    V_{\rm eff_{\rm corr}}^{n=1}(x)&=V_{S_{\rm corr}}^{n=1}(x)-V_{C_{\rm corr}}^{n=1}(x)\nonumber\\
    &=-2\pi G\rho r_0\Bigg[R+\frac{R^2-x^2}{2x}\ln\frac{R+x}{R-x}-(R-x)\ln\frac{\sqrt{(R-x)^2+a^2}}{R-x}\nonumber\\
&\quad-(R+x)\ln\frac{\sqrt{(R+x)^2+a^2}}{R+x}-a\tan^{-1}\!\left(\frac{R-x}{a}\right)-a\tan^{-1}\!\left(\frac{R+x}{a}\right)\!\Bigg].
\end{align}

The gravitational potential at any distance $x$ from the origin outside a full sphere, based on the power law formula with $n=1$ can be found by adding the Newtonian part $V_S^N(x)$ and the deviation part $V_S^{n=1}(x)$. As we saw above, the gravitational potential at a distance $x$ along the axis of symmetry of a uniform disk of radius $y$ and surface mass density $\sigma$ for $n=1$ is $-2\pi G\sigma r_0\ln(\sqrt{x^2+y^2}/x)$. By integrating this potential over the region $r\in[0,R]$, which lies to the left and to the right of the center of the sphere, and using $y^2=R^2-r^2$, where $r$ is the distance of the disk of radius $y$ from the center of the sphere, we find the following correction to the Newtonian potential outside the sphere,
\begin{align}\label{OutsideSphereYuk3}
V_{S_{\rm corr}}^{n=1(\rm Out)}(x)&=-2\pi G\rho r_0\int_0^R\left(\ln\frac{\sqrt{x^2+R^2+2xr}}{x+r}+\ln\frac{\sqrt{x^2+R^2-2xr}}{x-r}\right){\rm d}r\nonumber\\
&=-\pi G\rho r_0\left[\frac{(R+x)^2}{x}\ln(x+R)-\frac{(R-x)^2}{x}\ln(x-R)-2R\right]\nonumber\\
&\quad+2\pi G\rho r_0\int_0^R\left[\ln(x+r)+\ln(x-r)\right]{\rm d}r.
\end{align}
Similarly, recalling that the tunnel inside the sphere has the radius $a$ and the length $2R$, we would have the following gravitational potential outside the missing cylinder of radius $a$ and length $2R$:
\begin{align}\label{OutsideCylinderYuk}
V_{C_{\rm corr}}^{n=1(\rm Out)}(x)&=-2\pi G\rho r_0\int_0^R\left(\ln\frac{\sqrt{(x+r)^2+a^2}}{x+r}+\ln\frac{\sqrt{(x-r)^2+a^2}}{x-r}\right){\rm d}r\nonumber\\
&=-\pi G\rho r_0\Bigg[(x+R)\ln\left[(x+R)^2+a^2\right]-(x-R)\ln\left[(x-R)^2+a^2\right]\nonumber\\
&\quad-4R+2a\tan^{-1}\left(\frac{x+R}{a}\right)-2a\tan^{-1}\left(\frac{x-R}{a}\right)\Bigg]\nonumber\\
&\quad+2\pi G\rho r_0\int_0^R\left[\ln(x+r)+\ln(x-r)\right]{\rm d}r.
\end{align}
Therefore, the effective gravitational potential outside the sphere with the tunnel drilled inside it is $V_{{\rm eff}_{\rm Corr}}^{n=1(\rm Out)}(x)=V_{S_{\rm corr}}^{n=1(\rm Out)}(x)-V_{C_{\rm corr}}^{n=1(\rm Out)}(x)$, which yields, 
\begin{equation}\label{OutsideSphereYuk4}
V_{{\rm eff}_{\rm Corr}}^{n=1(\rm Out)}(x)\approx-2\pi G\rho r_0\left(\frac{x^2-R^2}{2x}\ln\frac{x-R}{x+R}+R\right).
\end{equation}
We have dropped terms of order $\mathcal{O}(a)$ and higher since they only bring a correction of the order $\mathcal{O}(ar_0)$. The extra induced phase shift due to the trip made by the particle outside the sphere is then:
\begin{align}\label{OutsidePhaseYuk}
    \Delta\phi_{\rm Corr}^{n=1(\rm Out)}&=-\frac{m^2\lambda_0}{\pi\hbar^2}\int_R^{L}V_{{\rm eff}_{\rm Corr}}^{n=1(\rm Out)}(x){\rm d}x\nonumber\\
    &=\frac{3GMm^2\lambda_0r_0}{4\pi\hbar^2R}\Bigg[\Re\left({\rm Li}_2\left(\frac{L}{R}\right)\right)\!-\!{\rm Li}_2\left(-\frac{L}{R}\right)\!-\!\frac{\pi^2}{4}\!+\!\frac{L^2-R^2}{2R^2}\ln\frac{L-R}{L+R}\!+\!\frac{L-R}{R}\Bigg].
\end{align}
Here, ${\rm Li}_2(z)$ represents Spence's (or dilogarithm) function \cite{FormulasBook} and $\Re\left({\rm Li}_2(z)\right)$ means that we take the real part of the function. It is clear from this expression that when the reflector and the beam splitter are not very far from the sphere, this extra phase shift correction remains negligible compared the phase shift (\ref{PhasePL1}) caused by the motion of the particle inside the tunnel.
\subsection{Case: $n=2$.}
We saw that for the case $n=2$, the correction to the Newtonian gravitational potential due to a disk of radius $y$ and mass density $\sigma$, at a distance $x$ along the axis of symmetry, is  $-2\pi G\sigma r_0^2[x^{-1}-(x^2+y^2)^{-1/2}]$. Therefore, by setting now, $g(s)=1/s-1/\sqrt{s^2+a^2}$, we find the following correction $V_{C_{\rm corr}}^{n=2}(x)$ to the Newtonian potential inside a full cylinder of radius $a$ and length $2R$,
\begin{align}\label{CylinderPLn2}
V_{C_{\rm corr}}^{n=2}(x)&=-2\pi G\rho r_0^2\left(\int_0^{R-x}+\int_0^{R+x}\right)g(s)\,{\rm d}s\nonumber\\
&=-2\pi G\rho r_0^2\Bigg[\int_0^{R-x}{\rm d}s/s+\int_0^{R+x}{\rm d}s/s-\ln\frac{R-x+\sqrt{(R-x)^2+a^2}}{a}\nonumber\\
&\quad-\ln\frac{R+x+\sqrt{(R+x)^2+a^2}}{a}\Bigg].
\end{align}
The first two integrals in the second line are both divergent but, as we shall see shortly, these two integrals do not contribute to the effective potential because they cancel exactly with a similar contribution from the full sphere.

In order to compute the correction $V_{S_{\rm corr}}^{n=2}(x)$ to the potential inside a full sphere, we shall use for convenience the function $g(s,y)=1/s-1/\sqrt{s^2+y^2}$. For the integration, we take into account as usual the shape of the sphere by using the fact that $y^2=R^2-(s+x)^2$ for the disks on the left of the point $x$ while $y^2=R^2-(s-x)^2$ to the right of the point $x$. Thus, we find,
\begin{align}\label{SpherePLn2}
V_{S_{\rm corr}}^{n=2}(x)&=-2\pi G\rho r_0^2\left(\int_0^{R-x}+\int_0^{R+x}\right)g(s,y)\,{\rm d}s\nonumber\\
&=-2\pi G\rho r_0^2\left[\int_0^{R-x}\!{\rm d}s/s+\int_0^{R+x}\!{\rm d}s/s\!-\!2\right].
\end{align}
Combining these corrections to the potential inside the sphere and the cylinder with the Newtonian potential inside each one of the latter, found in \ref{A}, we compute the effective potential inside the tunnel for case $n=2$ to be,
\begin{align}\label{VeffPLn2}
    V_{\rm eff_{\rm corr}}^{n=2}(x)&=V_{S_{\rm corr}}^{n=2}(x)-V_{C_{\rm corr}}^{n=2}(x)\nonumber\\
    &=2\pi G\rho r_0^2\Bigg[2\!-\!\ln\frac{R+x+\sqrt{(R+x)^2+a^2}}{a}\!-\!\ln\frac{R-x+\sqrt{(R-x)^2+a^2}}{a}\Bigg].
\end{align}

The gravitational potential at any distance $x$ from the origin outside a full sphere, based on the power law formula with $n=2$ can be found by adding the Newtonian part $V_S^N(x)$ and the deviation part $V_S^{n=2}(x)$. As we saw above, the gravitational potential at a distance $x$ along the axis of symmetry of a uniform disk of radius $y$ and surface mass density $\sigma$ for $n=2$ is $-2\pi G\sigma r_0^2[x^{-1}-(x^2+y^2)^{-1/2}]$. By integrating this potential over the region $r\in[0,R]$, which lies to the left and right of the center of the sphere, and using $y^2=R^2-r^2$, where $r$ is the distance of the disk of radius $y$ from the center of the sphere, we find,
\begin{align}\label{OutsideSphereYuk5}
V_{S_{\rm corr}}^{n=2(\rm Out)}(x)&=-2\pi G\rho r_0^2\int_0^R\Bigg(\frac{1}{x+r}+\frac{1}{x-r}-\frac{1}{\sqrt{x^2+R^2+2xr}}-\frac{1}{\sqrt{x^2+R^2-2xr}}\Bigg){\rm d}r\nonumber\\
&=4\pi G\rho r_0^2-2\pi G\rho r_0^2\int_0^R\left(\frac{1}{x+r}+\frac{1}{x-r}\right){\rm d}r.
\end{align}
Because of the tunnel inside the sphere, we would have the following gravitational potential outside a missing cylinder of radius $a$ and length $2R$:
\begin{align}\label{OutsideCylinderYuk}
V_{C_{\rm corr}}^{n=2(\rm Out)}(x)&=-2\pi G\rho r_0^2\int_0^R\Bigg(\frac{1}{x+r}+\frac{1}{x-r}-\frac{1}{\sqrt{(x+r)^2+a^2}}-\frac{1}{\sqrt{(x-r)^2+a^2}}\Bigg){\rm d}r\nonumber\\
&=-2\pi G\rho r_0^2\ln\frac{\sqrt{(x-R)^2+a^2}+x-R}{\sqrt{(x+R)^2+a^2}+x+R}-2\pi G\rho r_0^2\int_0^R\left(\frac{1}{x+r}+\frac{1}{x-r}\right){\rm d}r.
\end{align}
Therefore, the effective gravitational potential outside the sphere with the tunnel drilled inside it is $V_{{\rm eff}_{\rm Corr}}^{n=2(\rm Out)}(x)=V_{S_{\rm corr}}^{n=2(\rm Out)}(x)-V_{C_{\rm corr}}^{n=2(\rm Out)}(x)$, which gives, 
\begin{equation}\label{OutsideSphereYuk6}
V_{{\rm eff}_{\rm Corr}}^{n=2(\rm Out)}(x)=2\pi G\rho r_0^2\left(2+\ln\frac{x-R}{x+R}\right).
\end{equation}
Here, we have again dropped terms quadratic in $a$. The extra induced phase shift due to the trip made by the particle outside the sphere is then:
\begin{align}\label{OutsidePhaseN=2}
    \Delta\phi_{\rm Corr}^{n=2(\rm Out)}&=-\frac{m^2\lambda_0}{\pi\hbar^2}\int_R^{L}V_{{\rm eff}_{\rm Corr}}^{n=2(\rm Out)}(x){\rm d}x\nonumber\\
    &\approx\frac{3GMm^2\lambda_0r_0^2}{2\pi\hbar^2R^3}\left[R\ln\frac{L^2-R^2}{R^2}-L\ln\frac{L-R}{L+R}+2(R-L)-R\ln4\right].
\end{align}
For $L$ not very different from $R$, this additional phase shift is again negligible compared to the phase shift correction (\ref{PhasePL2}) acquired by the particle through its journey along the tunnel.
\subsection{Case: $n=3$.}
We saw that for the case $n=3$, the correction to the Newtonian gravitational potential due to a disk of radius $y$ and mass density $\sigma$, at a distance $x$ along the axis of symmetry, is  $-\pi G\sigma r_0^3[1/x^{2}-1/(x^2+y^2)]$. Therefore, by setting now, $g(s)=1/s^{2}-1/(s^2+a^2)$, we find the following correction $V_{C_{\rm corr}}^{n=3}(x)$ to the Newtonian potential inside a full cylinder of radius $a$ and length $2R$,
\begin{align}\label{CylinderPLn3}
V_{C_{\rm corr}}^{n=3}(x)&=-\pi G\rho r_0^3\left(\int_0^{R-x}+\int_0^{R+x}\right)g(s)\,{\rm d}s\nonumber\\
&=-\pi G\rho r_0^3\Bigg[\int_0^{R-x}{\rm d}s/s^2+\int_0^{R+x}{\rm d}s/s^2-\frac{1}{a}\tan^{-1}\left(\frac{R-x}{a}\right)-\frac{1}{a}\tan^{-1}\left(\frac{R+x}{a}\right)\Bigg].
\end{align}
The first two integrals in the second line are both divergent but, again, as we shall see shortly, these two integrals do not contribute to the effective potential because they cancel exactly with a similar contribution from the full sphere.

In order to compute the correction $V_{S_{\rm corr}}^{n=3}(x)$ to the potential inside a full sphere, we shall use for convenience the function $g(s,y)=1/s^{2}-1/(s^2+y^2)$. For the integration, we should take into account again the shape of the sphere by using the fact that $y^2=R^2-(s-x)^2$ for the disks on the right of the point $x$ while $y^2=R^2-(s+x)^2$ to the left of the point $x$. Thus we find,
\begin{align}\label{SpherePLn3}
V_{S_{\rm corr}}^{n=3}(x)&=-\pi G\rho r_0^3\left(\int_0^{R-x}+\int_0^{R+x}\right)g(s,y)\,{\rm d}s\nonumber\\
&=-\pi G\rho r_0^3\Bigg[\int_0^{R-x}{\rm d}s/s^2+\int_0^{R+x}{\rm d}s/s^2-\frac{1}{x}\ln\frac{R+x}{R-x}\Bigg].
\end{align}
Combining these corrections to the potential inside the sphere and the cylinder with the Newtonian potential inside each one of the latter, found in \ref{A}, we compute the effective potential inside the tunnel for case $n=3$ to be,
\begin{align}\label{VeffPLn3}
    V_{\rm eff_{\rm corr}}^{n=3}(x)&=V_{S_{\rm corr}}^{n=3}(x)-V_{C_{\rm corr}}^{n=3}(x)\nonumber\\
    &=\pi G\rho r_0^3\Bigg[\frac{1}{x}\ln\frac{R+x}{R-x}-\frac{1}{a}\tan^{-1}\left(\frac{R+x}{a}\right)+\frac{1}{a}\tan^{-1}\left(\frac{x-R}{a}\right)\Bigg].
\end{align}

The gravitational potential at any distance $x$ from the origin outside a full sphere, based on the power law formula with $n=3$ can be found by adding the Newtonian part $V_S^N(x)$ and the deviation part $V_S^{n=3}(x)$. As we saw above, the gravitational potential at a distance $x$ along the axis of symmetry of a uniform disk of radius $y$ and surface mass density $\sigma$ for $n=3$ is $-\pi G\sigma r_0^3[1/x^2-1/(x^2+y^2)]$. By integrating this potential over the region $r\in[0,R]$, which lies to the left and right of the center of the sphere, and using $y^2=R^2-r^2$, where $r$ is the distance of the disk of radius $y$ from the center of the sphere, we find,
\begin{align}\label{OutsideSphereYuk7}
V_{S_{\rm corr}}^{(n=3)\rm Out}(x)&=-\pi G\rho r_0^3\int_0^R\Bigg(\frac{1}{(x+r)^2}+\frac{1}{(x-r)^2}-\frac{1}{x^2+R^2+2xr}-\frac{1}{x^2+R^2-2xr}\Bigg){\rm d}r\nonumber\\
&=-\pi G\rho r_0^3\left(\frac{1}{x}\ln\frac{x-R}{x+R}\right)-\pi G\rho r_0^3\int_0^R\left[\frac{1}{(x+r)^2}+\frac{1}{(x-r)^2}\right]{\rm d}r.
\end{align}
Similarly, recalling that the tunnel inside the sphere has the radius $a$ and the length $2R$, we have the following gravitational potential outside a missing cylinder of radius $a$ and length $2R$:
\begin{align}\label{OutsideCylinderYuk}
V_{C_{\rm corr}}^{n=3(\rm Out)}(x)&=-\pi G\rho r_0^3\int_0^R\Bigg(\frac{1}{(x+r)^2}+\frac{1}{(x-r)^2}-\frac{1}{(x+r)^2+a^2}-\frac{1}{(x-r)^2+a^2}\Bigg){\rm d}r\nonumber\\
&=\frac{\pi G \rho r_0^3}{a}\left[\tan^{-1}\left(\frac{x+R}{a}\right)-\tan^{-1}\left(\frac{x-R}{a}\right)\right]\qquad\nonumber\\
&\quad-\pi G\rho r_0^3\int_0^R\left[\frac{1}{(x+r)^2}+\frac{1}{(x-r)^2}\right]{\rm d}r.
\end{align}
Therefore, the effective gravitational potential outside the sphere with the tunnel drilled inside it is $V_{{\rm eff}_{\rm Corr}}^{n=3(\rm Out)}(x)=V_{S_{\rm corr}}^{n=3(\rm Out)}(x)-V_{C_{\rm corr}}^{n=3(\rm Out)}(x)$, which gives, 
\begin{equation}\label{OutsideSphereYuk8}
V_{{\rm eff}_{\rm Corr}}^{n=3(\rm Out)}(x)=-\frac{3GM r_0^3}{4R^3}\left[\frac{1}{x}\ln\frac{x-R}{x+R}+\frac{1}{a}\tan^{-1}\left(\frac{x+R}{a}\right)-\frac{1}{a}\tan^{-1}\left(\frac{x-R}{a}\right)\right].
\end{equation}
This effective potential is the same as the one found in the interior of the sphere at a distance $x$ from the center of the latter. The extra induced phase shift due to the trip made by the particle outside the sphere is then,
\begin{align}\label{OutsidePhaseN=3}
    \Delta\phi_{\rm Corr}^{n=3(\rm Out)}&=-\frac{m^2\lambda_0}{\pi\hbar^2}\int_R^{L}V_{{\rm eff}_{\rm Corr}}^{n=3(\rm Out)}(x){\rm d}x\nonumber\\
    &\approx\frac{3GMm^2\lambda_0r_0^2}{4\pi\hbar^2aR^3}\Bigg[(L+R)\tan^{-1}\left(\frac{L+R}{a}\right)-(L-R)\tan^{-1}\left(\frac{L-R}{a}\right)\nonumber\\
    &\quad-2R\tan^{-1}\left(\frac{2R}{a}\right)\Bigg].
\end{align}
We have kept here only the leading terms which are inversely proportional to $a$. For $L$ not very far off from $R$, this extra phase shift is negligible (and even more so for very small $a$) compared to the phase shift correction (\ref{PhasePL3}) due to the journey of the particle inside the sphere.
\subsection{Case: $n=4$.}
We saw that for the case $n=4$, the correction to the Newtonian gravitational potential due to a disk of radius $y$ and mass density $\sigma$, at a distance $x$ along the axis of symmetry, is  $-\frac{2}{3}\pi G\sigma r_0^4[1/x^{3}-1/(x^2+y^2)^{3/2}]$. Therefore, by setting now, $g(s)=1/s^{3}-1/(s^2+a^2)^{3/2}$, we find the following correction $V_{C_{\rm corr}}^{n=4}(x)$ to the Newtonian potential inside a full cylinder of radius $a$ and length $2R$,
\begin{align}\label{CylinderPLn4}
V_{C_{\rm corr}}^{n=4}(x)&=-\frac{2\pi}{3} G\rho r_0^4\left(\int_0^{R-x}+\int_0^{R+x}\right)g(s)\,{\rm d}s\nonumber\\
&\quad=-\frac{2\pi}{3} G\rho r_0^4\Bigg[\int_0^{R-x}{\rm d}s/s^3+\int_0^{R+x}{\rm d}s/s^3-\frac{R-x}{a^2\sqrt{(R-x)^2+a^2}}\nonumber\\
&\qquad-\frac{R+x}{a^2\sqrt{(R+x)^2+a^2}}\Bigg].
\end{align}
Again, the first two integrals in the second line are both divergent but, as with the previous cases we saw above, these two integrals do not contribute to the effective potential because they cancel exactly with a similar contribution from the full sphere.

In order to compute the correction $V_{S_{\rm corr}}^{n=4}(x)$ to the potential inside a full sphere, we shall use for convenience the function $g(s,y)=1/s^{3}-1/(s^2+y^2)^{3/2}$. For the integration, we should take into account again the shape of the sphere by using the fact that $y^2=R^2-(s-x)^2$ for the disks on the right of the point $x$ while $y^2=R^2-(s+x)^2$ to the left of the point $x$. Thus we find,
\begin{align}\label{SpherePLn4}
V_{S_{\rm corr}}^{n=4}(x)&=-\frac{2\pi}{3}G\rho r_0^4\left(\int_0^{R-x}+\int_0^{R+x}\right)g(s,y)\,{\rm d}s\nonumber\\
&=-\frac{2\pi}{3} G\rho r_0^4\Bigg[\int_0^{R-x}{\rm d}s/s^3+\int_0^{R+x}{\rm d}s/s^3-\frac{2}{R^2-x^2}\Bigg].
\end{align}
Combining these corrections to the potential inside the sphere and the cylinder with the Newtonian potential inside each one of the latter, found in \ref{A}, we compute the effective potential inside the tunnel for case $n=4$ to be,
\begin{align}\label{VeffPLn4}
    V_{\rm eff_{\rm corr}}^{n=4}(x)&=V_{S_{\rm corr}}^{n=4}(x)-V_{C_{\rm corr}}^{n=4}(x)\nonumber\\
    &=\frac{2\pi}{3} G\rho r_0^4\Bigg[\frac{2}{R^2-x^2}-\frac{R-x}{a^2\sqrt{(R-x)^2+a^2}}-\frac{R+x}{a^2\sqrt{(R+x)^2+a^2}}\Bigg].
\end{align}
\subsection{Case: $n=5$.}
We saw that for the case $n=5$, the correction to the Newtonian gravitational potential due to a disk of radius $y$ and mass density $\sigma$, at a distance $x$ along the axis of symmetry, is  $-\frac{\pi}{2} G\sigma r_0^5[1/x^{4}-1/(x^2+y^2)^{2}]$. Therefore, by setting now, $g(s)=1/s^{4}-1/(s^2+a^2)^{2}$, we find the following correction $V_{C_{\rm corr}}^{n=5}(x)$ to the Newtonian potential inside a full cylinder of radius $a$ and length $2R$,
\begin{align}\label{CylinderPLn5}
V_{C_{\rm corr}}^{n=5}(x)&=-\frac{\pi}{2} G\rho r_0^5\left(\int_0^{R-x}+\int_0^{R+x}\right)g(s)\,{\rm d}s\nonumber\\
&=-\frac{\pi}{2} G\rho r_0^5\Bigg[\int_0^{R-x}{\rm d}s/s^4+\int_0^{R+x}{\rm d}s/s^4-\frac{R-x}{2a^2((R-x)^2+a^2)}\nonumber\\
&\quad-\frac{R+x}{2a^2((R+x)^2+a^2)}-\frac{1}{2a^3}\tan^{-1}\left(\frac{R-x}{a}\right)-\frac{1}{2a^3}\tan^{-1}\left(\frac{R+x}{a}\right)\Bigg].
\end{align}
As usual, the first two integrals in the second line are both divergent but, again, as with the previous cases we saw above, these two integrals do not contribute to the effective potential because they cancel exactly with a similar contribution from the full sphere.

In order to compute the correction $V_{S_{\rm corr}}^{n=5}(x)$ to the potential inside a full sphere, we shall use for convenience the function $g(s,y)=1/s^{4}-1/(s^2+y^2)^{2}$. For the integration, we should take into account again the shape of the sphere by using the fact that $y^2=R^2-(s-x)^2$ for the disks on the right of the point $x$ while $y^2=R^2-(s+x)^2$ to the left of the point $x$. Thus we find,
\begin{align}\label{SpherePLn5}
V_{S_{\rm corr}}^{n=5}(x)&=-\frac{\pi}{2}G\rho r_0^5\left(\int_0^{R-x}+\int_0^{R+x}\right)g(s,y)\,{\rm d}s\nonumber\\
&=-\frac{\pi}{2} G\rho r_0^5\Bigg[\int_0^{R-x}{\rm d}s/s^4+\int_0^{R+x}{\rm d}s/s^4-\frac{2R}{(R^2-x^2)^2}\Bigg].
\end{align}
Combining these corrections to the potential inside the sphere and the cylinder with the Newtonian potential inside each one of the latter, as found in \ref{A}, we easily compute the effective potential inside the tunnel for the case $n=5$ to be,
\begin{align}\label{VeffPLn5}
    V_{\rm eff_{\rm corr}}^{n=5}(x)&=V_{S_{\rm corr}}^{n=5}(x)-V_{C_{\rm corr}}^{n=5}(x)\nonumber\\
    &=-\frac{\pi}{2} G\rho r_0^5\Bigg[-\frac{2R}{(R^2-x^2)^2}+\frac{R-x}{2a^2((R-x)^2+a^2)}+\frac{R+x}{2a^2((R+x)^2+a^2)}\nonumber\\
    &\quad+\!\frac{1}{2a^3}\tan^{-1}\left(\frac{R-x}{a}\right)\!+\!\frac{1}{2a^3}\tan^{-1}\left(\frac{R+x}{a}\right)\Bigg].
\end{align}
\subsection{Case: $n=6$.}
We saw that for the case $n=6$, the correction to the Newtonian gravitational potential due to a disk of radius $y$ and mass density $\sigma$, at a distance $x$ along the axis of symmetry, is  $-\frac{2\pi}{5} G\sigma r_0^6[1/x^{5}-1/(x^2+y^2)^{5/2}]$. Therefore, by setting now, $g(s)=1/s^{5}-1/(s^2+a^2)^{5/2}$, we find the following correction $V_{C_{\rm corr}}^{n=6}(x)$ to the Newtonian potential inside a full cylinder of radius $a$ and length $2R$,
\begin{align}\label{CylinderPLn6}
V_{C_{\rm corr}}^{n=6}(x)&=-\frac{2\pi}{5} G\rho r_0^6\left(\int_0^{R-x}+\int_0^{R+x}\right)g(s)\,{\rm d}s\nonumber\\
&=-\frac{2\pi}{5} G\rho r_0^6\Bigg[\int_0^{R-x}{\rm d}s/s^5+\int_0^{R+x}{\rm d}s/s^5-\frac{(R-x)(3a^2+2(R-x)^2)}{3a^4(a^2+(R-x)^2)^{3/2}}\nonumber\\
&\quad-\frac{(R+x)(3a^2+2(R+x)^2)}{3a^4(a^2+(R+x)^2)^{3/2}}\Bigg].
\end{align}
Again, the first two integrals in the second line are both divergent but, as with the previous cases we saw above, these two integrals do not contribute to the effective potential because they cancel exactly with a similar contribution from the full sphere.

In order to compute the correction $V_{S_{\rm corr}}^{n=6}(x)$ to the potential inside a full sphere, we shall use for convenience the function $g(s,y)=1/s^{5}-1/(s^2+y^2)^{5/2}$. For the integration, we should take into account again the shape of the sphere by using the fact that $y^2=R^2-(s-x)^2$ for the disks on the right of the point $x$ while $y^2=R^2-(s+x)^2$ to the left of the point $x$. Thus we find,
\begin{align}\label{SpherePLn6}
V_{S_{\rm corr}}^{n=6}(x)&=-\frac{2\pi}{5}G\rho r_0^6\left(\int_0^{R+x}+\int_0^{R-x}\right)g(s,y)\,{\rm d}s\nonumber\\
&=-\frac{2\pi}{5} G\rho r_0^6\Bigg[\int_0^{R+x}{\rm d}s/s^5+\int_0^{R-x}{\rm d}s/s^5-\frac{(6R^2+2x^2)}{3(R^2-x^2)^3}\Bigg].
\end{align}
Combining these corrections to the potential inside the sphere and the cylinder with the Newtonian potential inside each one of the latter, found in \ref{A}, we compute the effective potential inside the tunnel for case $n=6$ to be,
\begin{align}\label{VeffPLn6}
    V_{\rm eff_{\rm corr}}^{n=6}(x)&=V_{S_{\rm corr}}^{n=6}(x)-V_{C_{\rm corr}}^{n=6}(x)\nonumber\\
    &=\frac{2\pi}{5}G\rho r_0^6\Bigg[\frac{(6R^2+2x^2)}{3(R^2-x^2)^3}-\frac{(R-x)(3a^2+2(R-x)^2)}{3a^4(a^2+(R-x)^2)^{3/2}}-\frac{(R+x)(3a^2+2(R+x)^2)}{3a^4(a^2+(R+x)^2)^{3/2}}\Bigg]
\end{align}
\section{The gravitational potential between two hemispheres based on formula (\ref{Yukawa})}\label{C}
We show here how to compute the gravitational potential $V_{H\rm eff}(x)$ between two hemispheres needed in Section~\ref{sec:VI}. For that purpose, we one needs only subtract the gravitational potential $V_D(x)$ due to a full uniform disk of radius $R$ and thickness $2a$ from the gravitational potential (\ref{SphereYuk}) of a full uniform sphere of radius $R$. In this appendix, we are going then to compute the potential due to a full disk. 

The gravitational potential at a distance $x$ from the center of a full disk of thickness $2a$, of radius $R$ and of mass density $\rho$, due to the Newtonian part in formula (\ref{Yukawa}), can be found by following the same strategy as the one adopted in Ref.~\cite{LassBlitzer}. We find,
\begin{align}\label{DiskNew}
    V_D^N(x)&=-4G\rho\int_0^\pi\int_0^{r(\phi)}\int_0^{a}\frac{r}{\sqrt{r^2+z^2}}\,{\rm d}\phi\,{\rm d}r\,{\rm d}z\nonumber\\
    &=-2G\rho\int_0^\pi\left[a\sqrt{r^2(\phi)+a^2}+r^2(\phi)\ln\left(\frac{\sqrt{r^2(\phi)+a^2}+a}{r(\phi)}\right)-a^2\right]{\rm d}\phi.
\end{align}
In the first line, we integrated, first, the potential due to the mass element $r\,{\rm d}\phi\,{\rm d}r\,{\rm d}z$ over the thickness $2a$. In the second line, we integrated over the rest of the disk from $r=0$ to $r=r(\phi)=x\cos\phi+\sqrt{R^2-x^2\sin^2\phi}$ \cite{LassBlitzer}. Now, the second term in the integrand in Eq.~(\ref{DiskNew}) does not admit any analytical expression. As such, one is tempted to perform a series expansion in $x$ on that term in order to be able to integrate it. Unfortunately, given that our goal is to be able later to integrate the potential $V_D^N(x)$ from $x=0$ to $x=R$, it does not help much to find a series expansion in $x$ that would terminate at some power of $x$. If a series expansion is to be used, it can only be a full series. Therefore, we shall instead seek a rough estimate of such an integral. Since the thickness of the disk is $2a\ll R$, it is clear that $r(\phi)$ inside the argument of the square root and the logarithm is mainly much larger than $a$, except for $x\sim R$. As a consequence, assuming $r(\phi)\gg a$ for all $x\lesssim R$ shall very well serve our purpose here for then the main contribution to the integral $\int_0^R V_D^N(x){\rm d}x$, which comes from the values $x\lesssim R$, would not be much affected. Thus, up to the second order in $a$, the first term inside the square brackets of the integral (\ref{DiskNew}) can be approximated by $ar(\phi)$ and the logarithm can be approximated by $a/r(\phi)$. In this case, Eq.~(\ref{DiskNew}) reduces to,
\begin{equation}
V_D^N(x)\approx-2G\rho\int_0^\pi\left[2ar(\phi)-a^2\right]{\rm d}\phi\approx-2G\rho\left[2aRE\left(\frac{x^2}{R^2}\right)-\pi a^2\right].
\end{equation}
Here, the function $E(z)$ is the so-called complete elliptic integral of the second kind \cite{FormulasBook}.
It admits a series representation as it is linked to the classical hypergeometric function through, $E(z)=\frac{\pi}{2}\,_2F_1\left(-\frac{1}{2},\frac{1}{2};1;z^2\right)$. Therefore, using the power expansion of the hypergeometric functions \cite{FormulasBook}, we find the following power-expansion for the potential $V_D^N(x)$,
\begin{align}\label{DiskNewExpansion}
V_D^N(x)&=-2\pi G\rho a R\sum_{n=0}^{\infty}\frac{\left(-\frac{1}{2}\right)_n\left(\frac{1}{2}\right)_n}{n!\,(1)_n}\frac{x^{2n}}{R^{2n}}+2\pi G\rho a^2.
\end{align}
The symbol $(x)_n$ stands for the product $x(x+1)(x+2)\ldots(x+n-1)$ and is called the Pochhammer symbol. By definition, $(x)_0=1$. Note that this is a full series in $x$ and not a truncation of a series. This potential can therefore be integrated exactly over the diameter of the disk to provide the phase shift $\Delta\phi$ for Section~\ref{sec:VI}. The integral of (\ref{DiskNewExpansion}) gives the following result,
\begin{align}\label{DiskNewIntegrated}
\int_0^RV_D^N(x)\,{\rm d}x&\approx-2\pi G\rho a R^2\sum_{n=0}^{\infty}\frac{\left(-\frac{1}{2}\right)_n\left(\frac{1}{2}\right)_n}{(2n+1)n!\,(1)_n}+2\pi G\rho a^2R\nonumber\\
&\equiv-2\pi G\rho aR\left(RI_D^N-a\right).
\end{align}
The factor $I_D^N$ represents the infinite sum. The number of terms of the sum to keep depends on the degree of precision one would wish to achieve which, in turn, depends on the order of magnitude of $a$.

Similarly, the gravitational potential inside the disk due to the Yukawa-like term in formula (\ref{Yukawa}) can be found as follows, 
\begin{align}\label{DiskYuk}
    V_D^Y(x)&=-4G\rho\alpha\int_0^\pi\int_0^{r(\phi)}\int_0^{a}\frac{re^{-\sqrt{r^2+z^2}/\lambda}}{\sqrt{r^2+z^2}}\,{\rm d}\phi\,{\rm d}r\,{\rm d}z\nonumber\\
    &\approx-4G\rho\alpha e^{-a/\lambda}\int_0^\pi\int_0^{r(\phi)}\int_0^{a}e^{-r/\lambda}\,{\rm d}\phi\,{\rm d}r\,{\rm d}z\nonumber\\
    &\approx-4G\lambda\rho\alpha ae^{-a/\lambda}\int_0^\pi \left(1-e^{-r(\phi)/\lambda}\right){\rm d}\phi\nonumber\\
    &\approx-4\pi G\lambda\rho\alpha ae^{-a/\lambda}\left(1-e^{-\frac{R-x}{\lambda}}\right).
\end{align}
Given that the integral in the first line does not admit any analytical expression, we have approximated it in the second line by noticing that the main contribution to the integral comes from $r(\phi)\gg a$. This allowed us to absorb the $z^2$ inside the square root. Consequently, we had then to take the lowest limit, $e^{-(a+r)/\lambda}$, of the exponential in the numerator. In fact, thanks to these approximations, the integral is easily evaluated in the third line. The latter expression does not admit any analytical expression either. For this reason, we had to take the lowest absolute value of $r(\phi)$, which is $R-x$. With such an approximate result, we find, up to the first order in $\lambda$ ($\ll R$) the following integral,
\begin{align}\label{DiskYukIntegrated}
\int_0^RV_D^Y(x)\,{\rm d}x\approx-4\pi G\lambda\rho R\alpha ae^{-a/\lambda}.
\end{align}


\end{document}